\documentclass[10pt,journal,twocolumn]{IEEEtran}
\usepackage{tabularx}
\usepackage{subcaption,wrapfig,graphicx,booktabs,fancyhdr,amsmath,amsfonts}
\usepackage{cite,bm,amsthm,url,multirow,times,enumitem,comment}
\usepackage{siunitx}
\usepackage{balance}
\newcommand{\vb}{\boldsymbol}
\newcommand{\vbh}[1]{\hat{\boldsymbol{#1}}}

\newcommand{\ba}{\begin{array}}
\newcommand{\ea}{\end{array}}
\newcommand{\sinc}{{\rm sinc}}

\newcommand{\cnr}{C/N_0}

\newcommand{\E}[1]{\mathbb{E}\left[ #1 \right]}

\DeclareMathAlphabet{\mathpzc}{OT1}{pzc}{m}{it}

\begin{document}
\title{First results from three years of GNSS Interference Monitoring from Low
  Earth Orbit \thanks{Correspondence: matthew.murrian@utexas.edu}}

\author{
  \IEEEauthorblockN{Matthew J. Murrian\IEEEauthorrefmark{1},
    Lakshay Narula\IEEEauthorrefmark{2}, Peter
    A. Iannucci\IEEEauthorrefmark{1},
    Scott Budzien\IEEEauthorrefmark{3},\\
    Brady W. O'Hanlon\IEEEauthorrefmark{4},
    Mark L. Psiaki\IEEEauthorrefmark{5},
    Todd E. Humphreys\IEEEauthorrefmark{1}} \\
  \IEEEauthorblockA{\IEEEauthorrefmark{1}\textit{Department of Aerospace
      Engineering and Engineering Mechanics, The University of Texas at
      Austin}} \\
  \IEEEauthorblockA{\IEEEauthorrefmark{2}\textit{Department of Electrical and Computer Engineering, The University of Texas at
      Austin}} \\
  \IEEEauthorblockA{\IEEEauthorrefmark{3}\textit{Naval Research Laboratory}} \\
  \IEEEauthorblockA{\IEEEauthorrefmark{4}\textit{Cornell University}} \\
  \IEEEauthorblockA{\IEEEauthorrefmark{5}\textit{Department of Aerospace and
      Ocean Engineering, Virginia Tech}}}
\maketitle

\begin{abstract}
  Observation of terrestrial GNSS interference (jamming and spoofing) from
  low-earth orbit (LEO) is a uniquely effective technique for characterizing
  the scope, strength, and structure of interference and for estimating
  transmitter locations.  Such details are useful for situational awareness,
  interference deterrence, and for developing interference-hardened GNSS
  receivers.  This paper presents the results of a three-year study of global
  interference, with emphasis on a particularly powerful interference source
  active in Syria since 2017.  It then explores the implications of such
  interference for GNSS receiver operation and design.
\end{abstract}


\pagestyle{plain}
\thispagestyle{empty} 


\section{Introduction}
This paper presents the results of a three-year study of terrestrial GNSS
interference as observed through a software-defined GNSS receiver operating
since February 2017 on the International Space Station (ISS).  The FOTON
receiver, developed by The University of Texas at Austin (UT) and Cornell
University, is part of a larger science experiment called GPS Radio Occultation
and Ultraviolet Photometry—Colocated (GROUP-C), an unclassified experiment
aboard the ISS that is part of the Space Test Program—Houston Payload 5
(STP-H5) payload. Serendipitous observations of GNSS interference in the
occultation data are an important early result of GROUP-C's scientific
objective to characterize GPS signals in the LEO environment.  This paper
discusses the interference signals detected, their effects, and interference
mitigation strategies for receivers deployed in LEO and terrestrial
environments.

The FOTON receiver is a science-grade spaceborne dual-frequency (GPS L1 and L2)
GNSS receiver \cite{lightsey2013demonstration}.  Three levels of FOTON data
are available for interference analysis: (1) raw 5.7 Msps intermediate
frequency (IF) samples output by the FOTON front-end's analog-to-digital
converter, (2) 100-Hz data-modulation-wiped complex correlation products, and
(3) 1-Hz standard GNSS observables (pseudorange, carrier phase,
carrier-to-noise ratio $\cnr$).

Although spaceborne GNSS sensors have been used for remote sensing via radio
occultation \cite{ao2009rising} and reflectometry \cite{jin2010gnss}, there
is little public literature exploring their use for monitoring terrestrial GNSS
interference.  \cite{isoz2014int} characterized interference observed at a LEO
satellite, and approximated the location for one source, but was mainly concerned
with determining whether the interference had a detrimental impact on GPS-RO
(radio occultation) meteorological products.  The more recent survey of GNSS
interference localization techniques in \cite{dempster2016interference} makes
no mention of single-receiver Doppler-based localization, whether space-based
or not.

General TDOA and FDOA interference localization has been extensively studied
\cite{ho1997geolocation,pattison2000sensitivity,griffin2002interferometric,amar2008localization,bhatti2015dissertation},
and such techniques have been applied for terrestrial interference localization
from geostationary orbit \cite{smith1989time,ho1993solution,
  haworth1997interference}.  Application of T/FDOA for localization from LEO
can be viewed as an extension of such demonstrations, with the lower-altitude
orbits enabling localization of much weaker signals.  Interference localization
using a single satellite has been explored in \cite{kalantari2016frequency},
but only simulation results are presented, and these unrealistically assume
perfect-tone interference with a known and constant frequency.

This paper makes three primary contributions. First, it improves on the global
survey technique in \cite{isoz2014int} by compensating for predictable $\cnr$
variations in the detection test.  Second, it presents the results of a
three-year study of global GNSS interference, with emphasis on a powerful
interference source active in Syria since 2017.  Via Doppler positioning using
the FOTON instrument on the ISS, an estimate of the transmitter's location is
obtained whose horizontal errors are less than 1 km with 99\% confidence based
on reasonable clock and noise models.  Such an accurate localization of a GNSS
interference source from LEO is without precedent in the open literature.
Third, this paper explores the implications of interference of the type
generated by the source active in Syria for GNSS receiver operation and design.

A preliminary version of this paper was published in \cite{murrian2019leoIon}.
The current version focuses on the observed interference, extends the analysis
period to June 2020, offers a more detailed analysis of localization accuracy,
and includes a new section exploring implications for GNSS receivers.

\section{Single-Satellite Terrestrial Source Geolocation}
\label{sec:single-satell-geol}
As a prelude to the presentation of results from the observation campaign, this
section introduces and analyzes the Doppler-based technique employed to
estimate the location of the interference source operating in Syria.

Assuming a carrier can be extracted from an interference signal,
single-satellite-based transmitter geolocation is possible from Doppler
measurements alone \cite{becker1992efficient, ellis2020use}.  The analysis
presented here emphasizes the effect of transmitter clock stability on
geolocation accuracy.

Consider a static transmitter emitting a signal at the GPS L1 frequency as
observed by a moving receiver.  Let $\lambda$ be the signal wavelength in
meters, $\vbh{r}$ the unit vector pointing from the transmitter to receiver,
expressed in Earth-centered-Earth-fixed (ECEF) coordinates, $\vb{v}_{\rm R}$
the receiver velocity with respect to the ECEF frame and expressed in ECEF in
m/s, and $\delta\dot{t}_{\rm R}$ the receiver clock frequency error in s/s, all
at the time of signal receipt. Further, let $\delta\dot{t}_{\rm T}$ be the
transmitter clock frequency error in s/s at the time of signal transmission,
and $w$ be a zero-mean Gaussian error term that models thermal noise,
ionospheric and tropospheric delay rates, and other minor effects, in Hz.  Then
the observed Doppler frequency in Hz at the receiver can be modeled as
\begin{equation}
\label{eq:doppler_meas_model}
f_{\rm D} =  -\vbh{r}^T \vb{v}_{\rm R}/\lambda -
c\left[\delta\dot{t}_{\rm R} - \delta\dot{t}_{\rm T}\left(1 -
    \delta\dot{t}_{\rm R}\right)\right]/\lambda + w
\end{equation}
where $c$ is the speed of light in m/s.  It is assumed that $\vb{v}_{\rm R}$,
$\delta\dot{t}_{\rm R}$, and the receiver position are known, e.g., via a GNSS
receiver co-located with the transmitted signal receiver.  The unknowns in
(\ref{eq:doppler_meas_model}) are transmitter position, which is embedded in
$\vbh{r}$, and $\delta\dot{t}_{\rm T}$.  The former is modeled as an unknown
constant and the latter as a random walk process that evolves as
\begin{equation}
  \label{eq:random_walk}
  \dot{\delta t}_{\rm T}(t_{k+1}) = \dot{\delta t}_{\rm T}(t_{k}) + v(t_k)
\end{equation}
Here, $v(t_k)$ is a discrete-time Gaussian random process with
$\E{v(t_k)} = 0$ and
$\E{v(t_k)v(t_j)} = 2 \pi^2 h_{-2} \delta t \delta_{k,j}, ~ \forall k,j$, where
$h_{-2}$ is the first parameter of the standard clock model based on the
fractional frequency error power spectrum, as given in
\cite[Chap. 8]{brown2012introKf}; $\delta t = t_{k+1} - t_k$ is the uniform
sampling interval; and $\delta_{k,j}$ is the Kronecker delta.

A transmitter could introduce any level of complexity to carrier-phase
frequency behavior; e.g., frequency modulation, frequency hopping, etc. Such
behaviors, if not discovered and appropriately modeled, would confound
single-pass geolocation efforts. Here, it is assumed that a nominally-constant
carrier frequency is intended by the transmitter and that it is operating in
steady-state conditions.  In fact, it will be assumed that $h_{-2}$ is
sufficiently small that $\dot{\delta t}_{\rm T}$ can be modeled as constant
over a short (e.g., 60-second) data capture interval.

Based on the above Doppler measurement model, a batch maximum likelihood
estimator \cite{crassidis2011optimal} can be developed to estimate the unknown
transmitter position and a constant value for $\dot{\delta t}_{\rm T}$ from a
collection of single-pass Doppler measurements.  If Doppler measurements from
multiple satellite passes are available, these can be combined for single-batch
estimation provided that a new value of $\dot{\delta t}_{\rm T}$ is estimated
for each pass. In other words, $\dot{\delta t}_{\rm T}$ is viewed as constant
over each short capture interval but variable from capture to capture.


When $\dot{\delta t}_{\rm T}$ is modeled as constant over a capture interval,
actual transmitter clock instability gives rise to Doppler measurement
errors. The impact of such errors on geolocation accuracy was analyzed via
Monte Carlo simulation for three levels of transmitter clock quality, from
a temperature-compensated crystal oscillator (TCXO) to a laboratory-grade
oven-controlled crystal oscillator (OCXO).  Simulation parameters were based
on the real-world interference capture discussed in the next section: the true
transmitter location was simulated to be $35.4$N latitude, $35.95$E longitude,
$48$m altitude; the receiver trajectory was taken from the ISS orbit during the
first 60 seconds of the capture interval on day 144 of 2018 (resulting in
$441.65$ km of total receiver displacement); and the measurement rate was $20$
Hz.  First, an error-free Doppler time history was generated based on this
scenario. Then, for each instance of the Monte Carlo simulation, an independent
realization of a Doppler error random process consistent with the clock model
being analyzed was generated and added to the error-free Doppler. Doppler error
was modeled as a random walk process consistent with (\ref{eq:random_walk}).  These
models assume a smooth compensation for temperature control, such as is common
for TCXOs used in GNSS receivers.  Additionally, $h_{-2}$ is assumed to dominate
frequency stability over each short capture interval.

$1000$ Monte Carlo trials were conducted for each of the three clock quality
levels.  Transmitter horizontal location estimation errors were observed to be
zero-mean and apparently Gaussian, and they were consistent with the formal
error ellipses of the associated linear least-squares estimator.  To determine
whether 1000 trials were sufficient for a confident error analysis, subgroups
of $250$ trials were randomly selected from the $1000$ trials and each of their
geolocation error ellipses were calculated.  Those subgroup samples were
observed to deviate less than $\pm10\%$ from the population mean with $99\%$
empirical confidence.  For example, $99\%$ of the subgroups for the TCXO
simulation had geolocation error ellipse estimates between $6900\pm660$
meters, semi-major axis, and $690\pm71$ meters, semi-minor axis.  Out of
\num{1e5} subgroup samples drawn for each clock quality level, none were
observed that deviated more than $17\%$ from the population mean. Thus,
1000-trial-based error ellipses for each clock quality level given in
Table~\ref{table:clock_quality_vs_geolocation_sigma} can be assumed to be no
more than $15\%$ smaller, on either axis, than the error ellipses that would be
produced in the limit of an infinite number of trials.

\begin{table}[htbp]
  \caption{Marginal contribution of transmitter frequency instability to
  a single-pass geolocation error ellipse.  The size of the $95\%$ horizontal
  geolocation error ellipse, in meters, is characterized by the semi-major ($a$)
  and semi-minor ($b$) axes.\label{table:clock_quality_vs_geolocation_sigma}}
  \centering
  \begin{tabular}{lcrl}
    \toprule
Clock Quality & $h_{-2}$ & $a$ (m) & $b$ (m) \\ \midrule
TCXO & \num{3e-21} & $6900$ & $690$ \\
Low-quality OCXO & \num{3e-23} & $720$ & $\num{72}$ \\
OCXO & \num{3e-25} & $67$ & $\num{7.4}$ \\ \bottomrule
\end{tabular}
\end{table}

Table~\ref{table:clock_quality_vs_geolocation_sigma} shows that marginal
contribution of transmitter frequency instability to single-pass geolocation
error grows precipitously with reduced transmitter clock quality.  These
results suggest that single-pass geolocation of a TCXO-based transmitter is
marginal at best, and could be even worse if the $h_{-2}$ values for TCXOs in
Table \ref{table:clock_quality_vs_geolocation_sigma} are optimistic. On the
other hand, if the transmitter is driven by an OCXO-quality clock, then clock
instability contributes less than 720 meters (low-quality OCXO) or 67 meters
(standard-quality OCXO).

The error ellipse characterized by $a$ and $b$ is highly eccentric, with
semi-minor axis oriented in the direction of satellite motion: e.g., if the
satellite is moving west to east then transmitter location will be best
resolved in that direction. It follows that additional satellite passes provide
the most benefit when, relative to the transmitter location, they are
geometrically dissimilar to previous passes.

\section{Analysis of Interference from Syria}
This section presents an in-depth analysis of a particular interference source
active on the east coast of the Mediterranean Sea during the period of this
paper's study, which spans from March 2017 to June 2020.  The analysis
illustrates the techniques that can be applied generally to study terrestrial
GNSS interference sources using signals collected in LEO.

Recording raw IF data in LEO and relaying these to the ground for processing is
an especially flexible approach well suited to studying new or
poorly-understood interference.  For the case presented here, the FOTON
receiver captured 1-minute intervals of raw 5.7-Msps two-bit-quantized IF
samples at the GPS L1 (1575.42 MHz) and the GPS L2 (1227.6 MHz) frequencies.
These data were packaged and downlinked via NASA's communications backbone.
Ground processing using the latest version of UT's software-defined GNSS
receiver \cite{humphreys2019deepUrbanIts} enabled analysis and tracking of all
radio frequency signals near GPS L1 and L2.

\begin{figure}
  \includegraphics[width=0.5\textwidth]{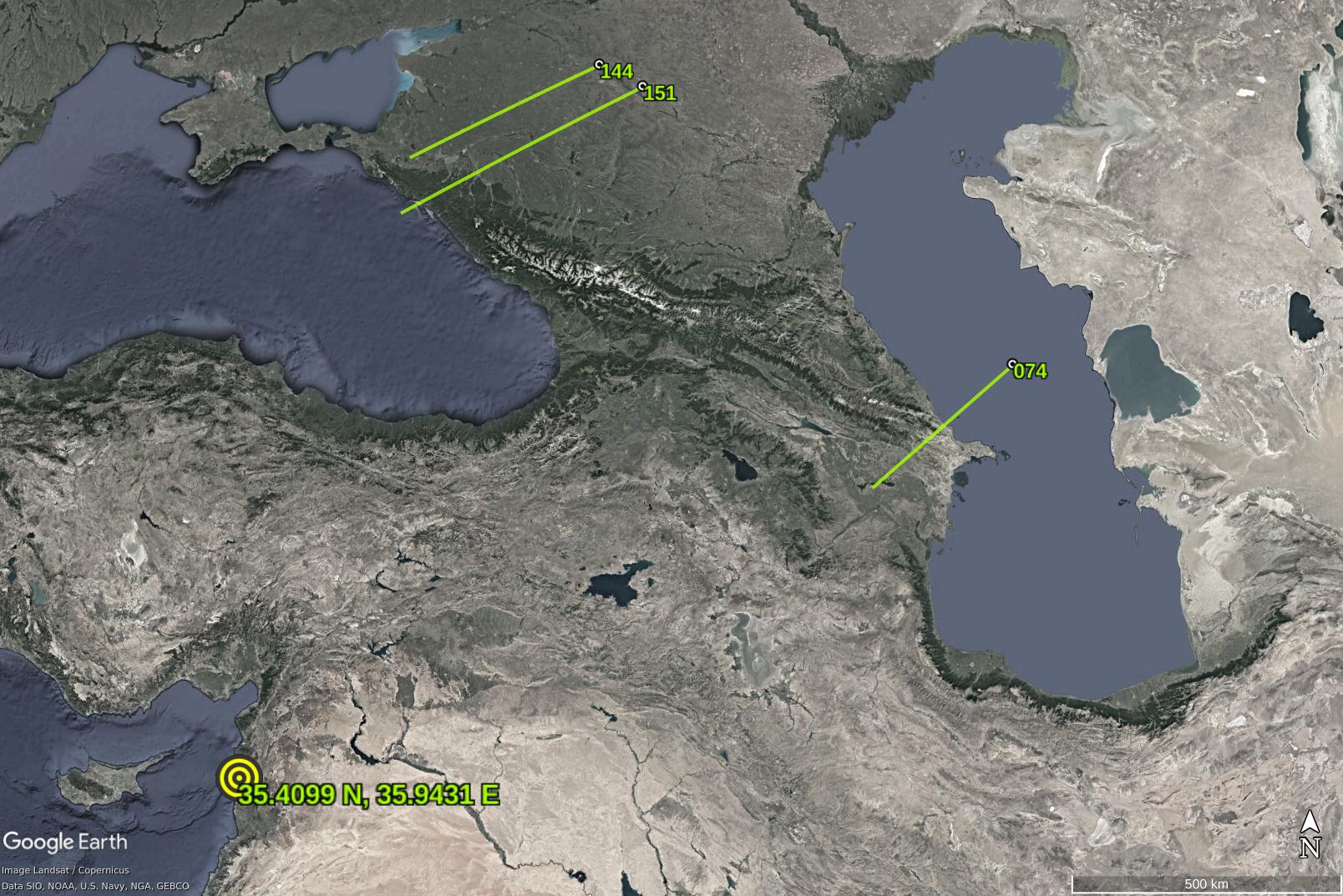}
  \caption{Ground tracks for interference-affected captures on days 74, 144,
    and 151 of 2018.  Each capture spans approximately 60 seconds.  The
    estimated transmitter location is marked on the west coast of Syria.}
  \label{fig:ground_tracks}
\end{figure}

The following observations are based on signals captured on three days in the
first half of 2018 along the ground tracks shown in
Fig. \ref{fig:ground_tracks}.

\subsection{Overview}
\label{sec:overview}
Strong interference is present in both the L1 and L2 bands, but the nature of
the interference is markedly different between the two bands.  At L2, the
interference is narrowband, whereas at L1 it is a wideband spread-spectrum
signal.  The L1 interference is a composite of individual signals with a common
carrier centered near GPS L1 but each having a unique GPS L1 C/A pseudo-random
number (PRN) spreading code.  Such interference can be categorized as
matched-code GNSS interference
\cite{humphreysGNSShandbook,psiakiNewBlueBookspoofing}.  Signals corresponding
to almost all GPS L1 C/A PRN codes from 1 to 32 have been detected.  When
tracked by the UT software-defined GNSS receiver, all false signals exhibit
$\cnr$ values greater than $40$ dB-Hz.  No discernible navigation data are
modulated on the false GPS L1 signals.  Moreover, the false signals are not
clean simulated GPS L1 C/A signals: they exhibit unexplained fading and
spectral characteristics.  No false Galileo BOC(1,1) signals were detected in
the L1 band.

The lack of navigation bit modulation renders the signals ineffective at
spoofing, but matched-code interference is a particularly potent form of
jamming \cite{humphreysGNSShandbook}.  Why different techniques were used at L1
and L2 is unknown.

While some authentic GPS L1 C/A signals in the data are effectively jammed, the
majority of authentic signals are still trackable owing to sufficient
separation of corresponding false and authentic signals in code-Doppler
space. Thus, a correct receiver navigation solution can still be formed despite
the interference.

\begin{figure*}
  \begin{subfigure}[b]{0.5\textwidth}
    \includegraphics[width=\textwidth]{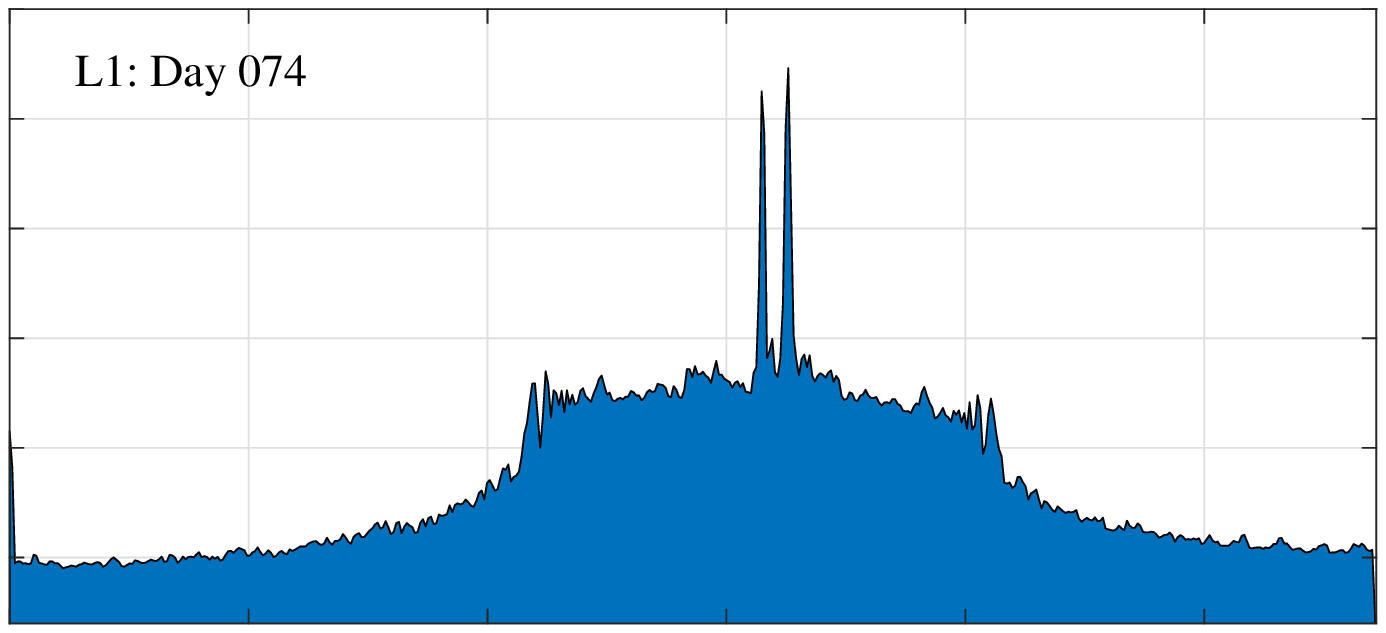}
  \end{subfigure}
  \begin{subfigure}[b]{0.5\textwidth}
    \includegraphics[width=\textwidth]{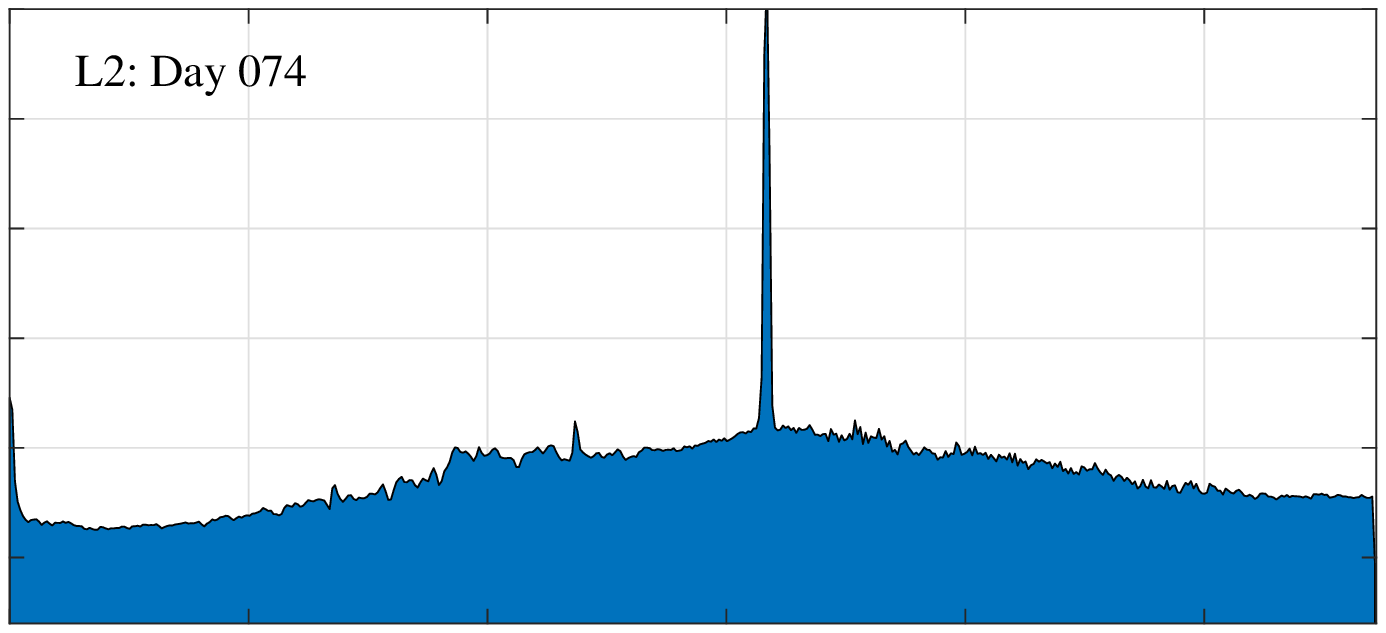}
  \end{subfigure}
  \begin{subfigure}[b]{0.5\textwidth}
    \includegraphics[width=\textwidth]{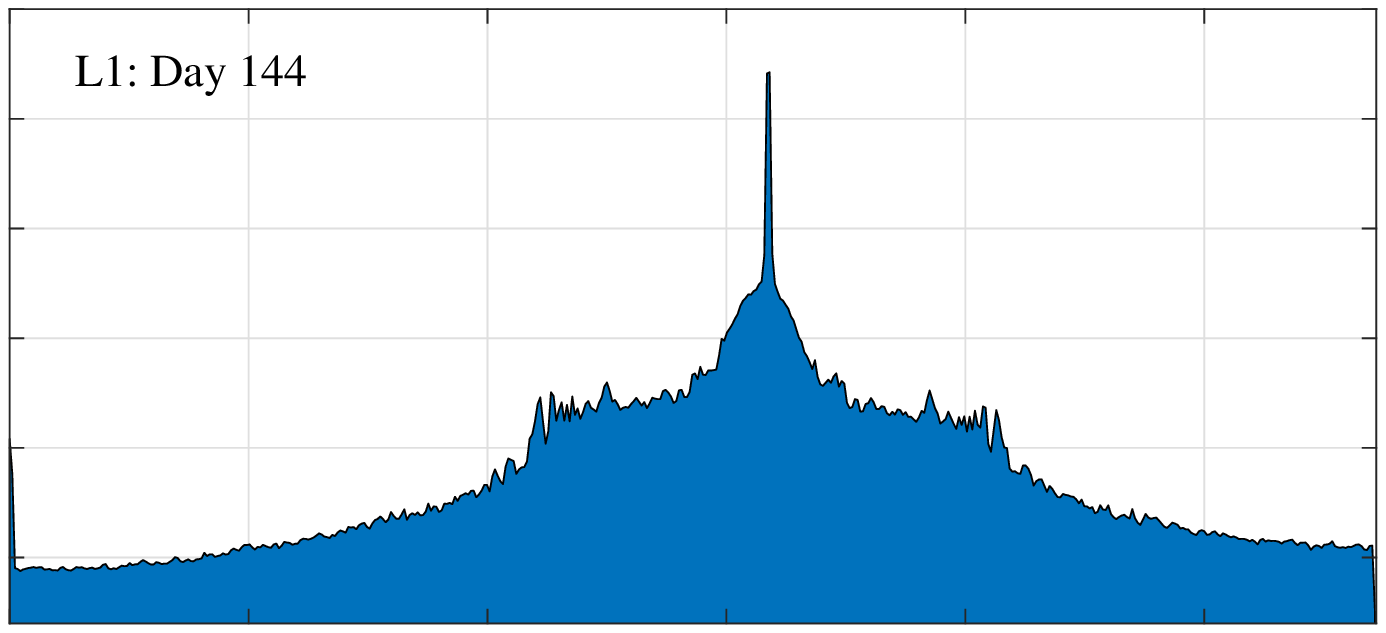}
  \end{subfigure}
  \begin{subfigure}[b]{0.5\textwidth}
    \includegraphics[width=\textwidth]{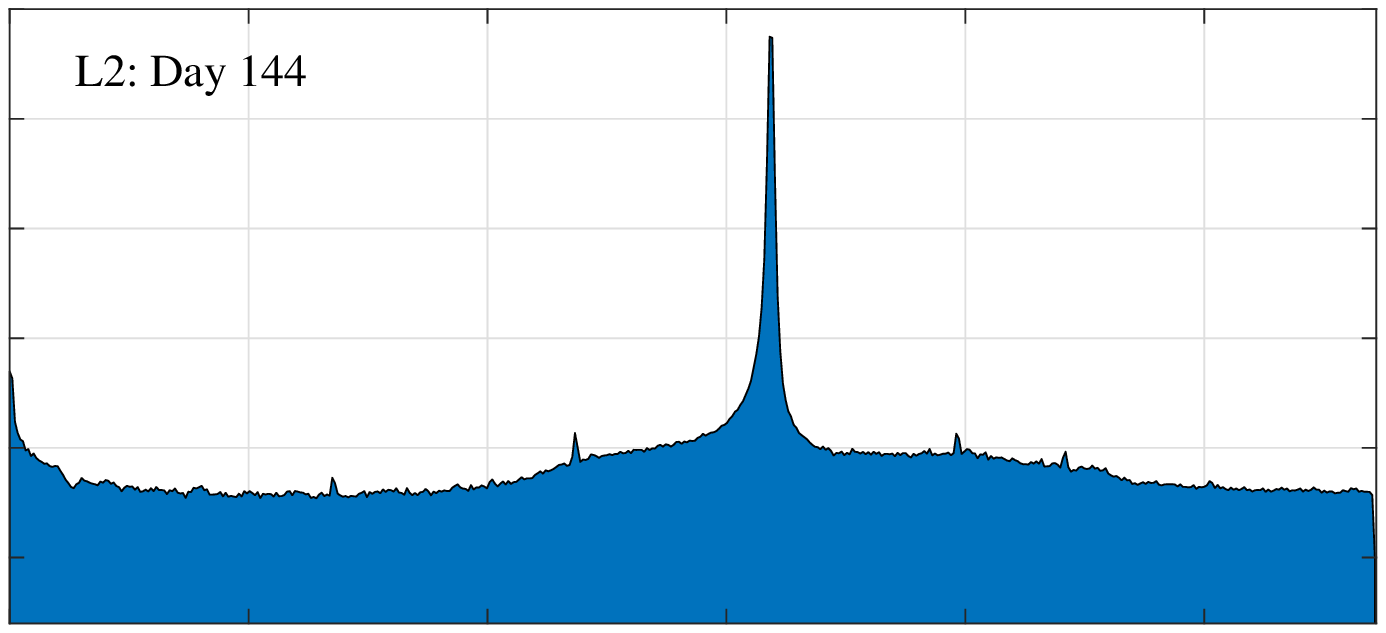}
  \end{subfigure}
   \begin{subfigure}[b]{0.5\textwidth}
    \includegraphics[width=\textwidth]{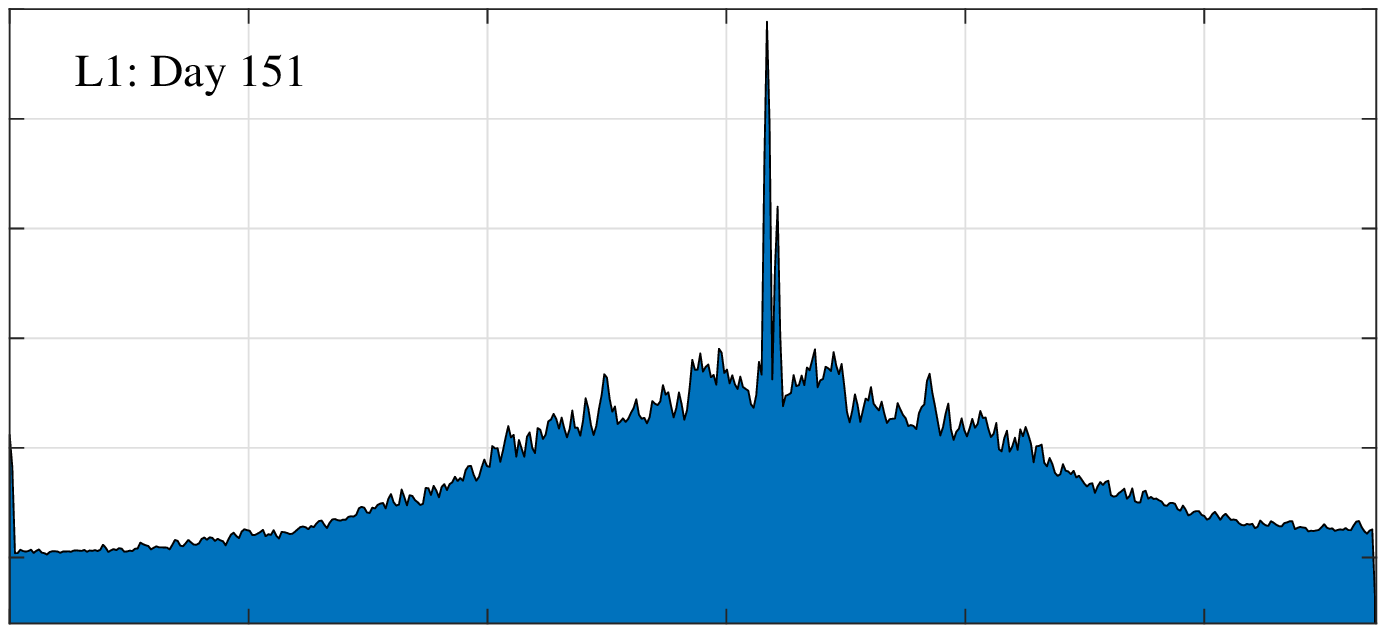}
  \end{subfigure}
  \begin{subfigure}[b]{0.5\textwidth}
    \includegraphics[width=\textwidth]{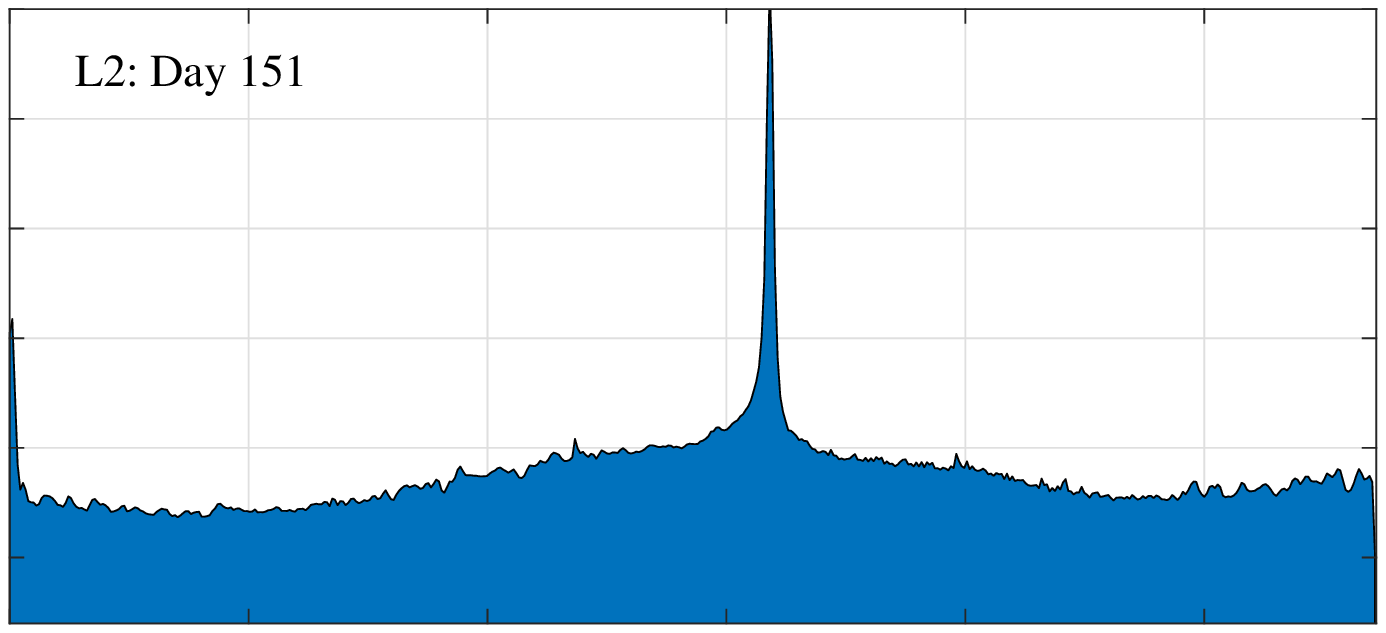}
  \end{subfigure}
   \begin{subfigure}[b]{0.5\textwidth}
    \includegraphics[width=\textwidth]{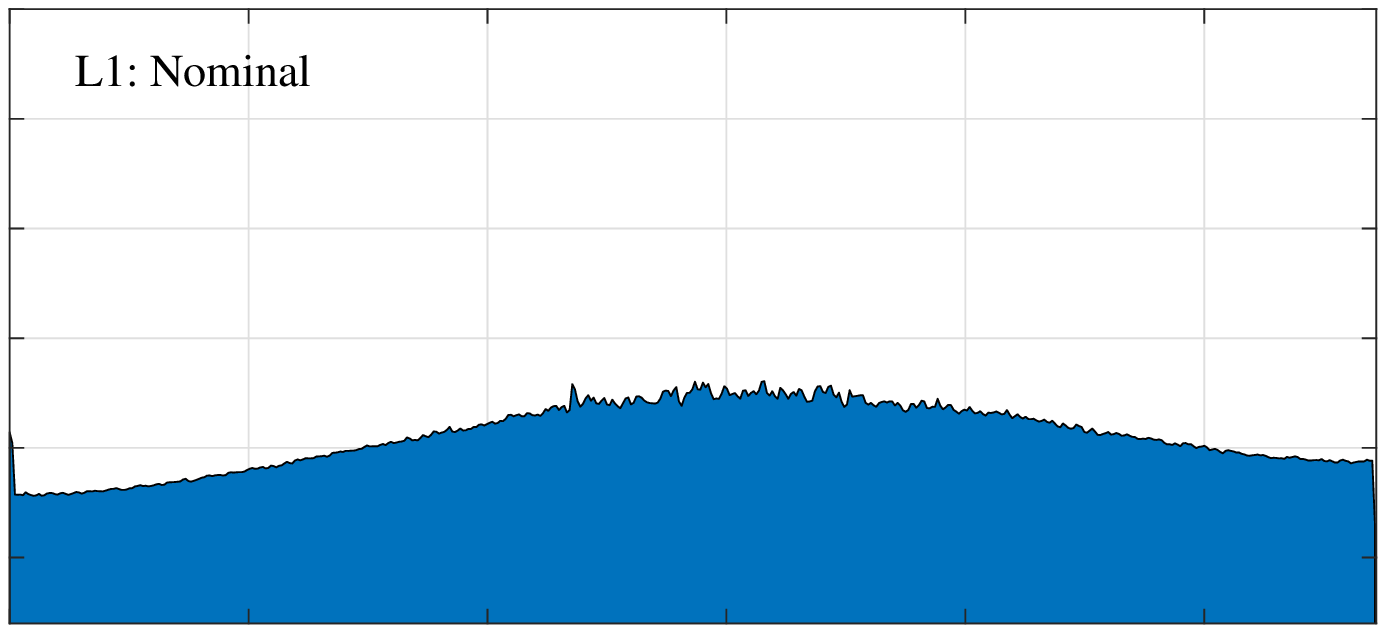}
  \end{subfigure}
  \begin{subfigure}[b]{0.5\textwidth}
    \includegraphics[width=\textwidth]{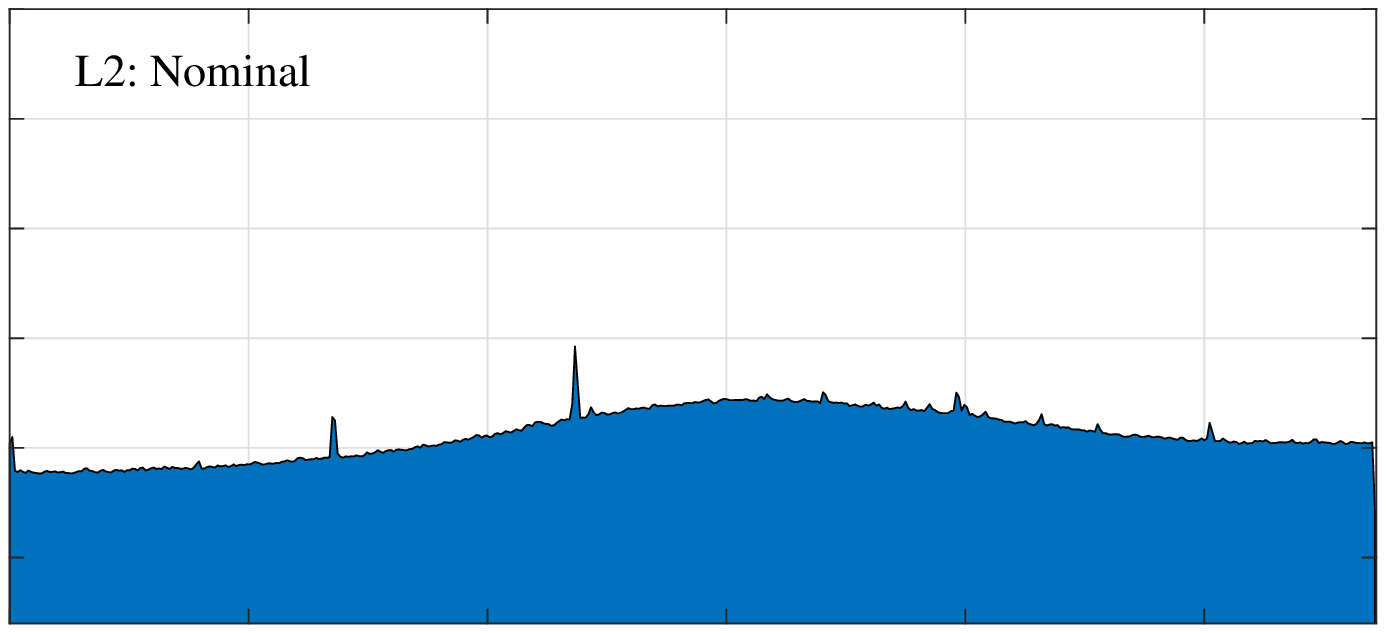}
  \end{subfigure}
  \caption{Power spectra centered near the GPS L1 (left column) and L2 (right
    column) frequencies from interference-affected data captured on days 74,
    144, and 151 of 2018 (top three rows), and from nominal data captured on
    day 158 of 2018 (bottom row).  The frequency span is approximately 3 MHz
    wide, scaled linearly with 0.5 MHz divisions.  All ordinate axes are in dB
    and scaled equivalently for ease of comparison.  Spectra are estimated by
    Welch's method \cite{welch1967use} from 1-second data intervals with a
    5.6-kHz frequency resolution.}
\label{fig:power_spectra}
\end{figure*}

\begin{figure}
  \begin{subfigure}[b]{0.5\textwidth}
    \includegraphics[width=\textwidth]{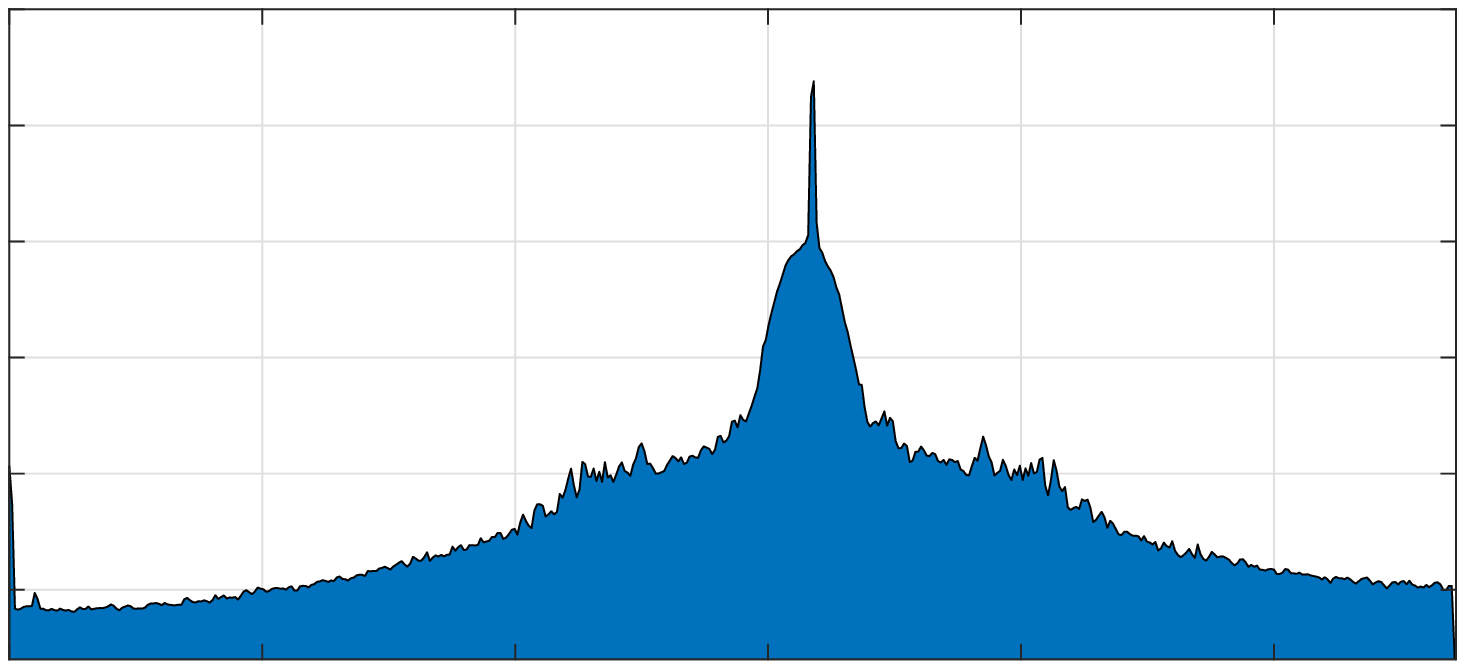}
  \end{subfigure}
  \begin{subfigure}[b]{0.5\textwidth}
    \includegraphics[width=\textwidth]{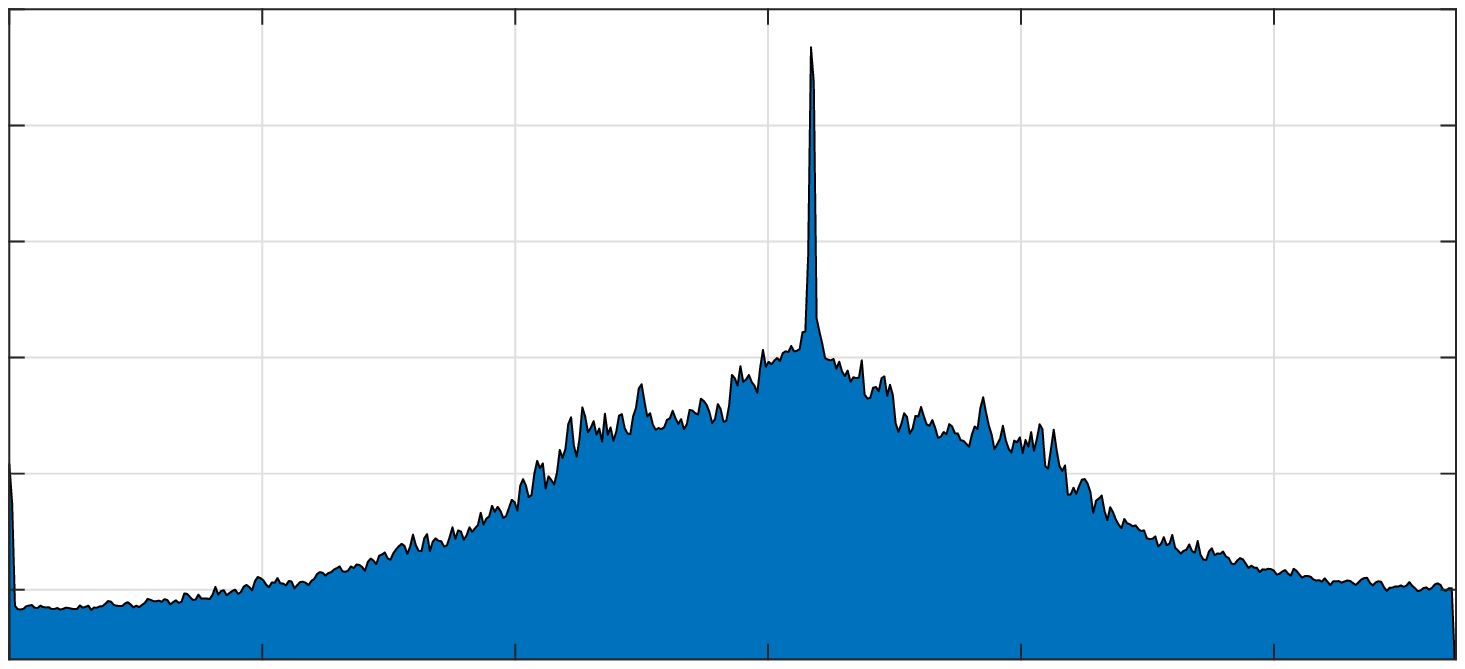}
  \end{subfigure}
  \caption{Power spectra near L1 for the day 144 capture showing maximum (top)
    and minimum (bottom) phases of the waxing and waning wideband
    ($\sim0.25$MHz) central interference prominence. The prominence oscillates
    with a period of approximately 5 seconds. The L1: Day 144 plot in
    Fig. \ref{fig:power_spectra} catches the prominence waning two seconds
    after the maximum shown in the top plot above.}
\label{fig:bump_wax_and_wane}
\end{figure}

\begin{figure}[!ht]
  \includegraphics[width=0.5\textwidth, trim=0 0 0 0,
  clip]{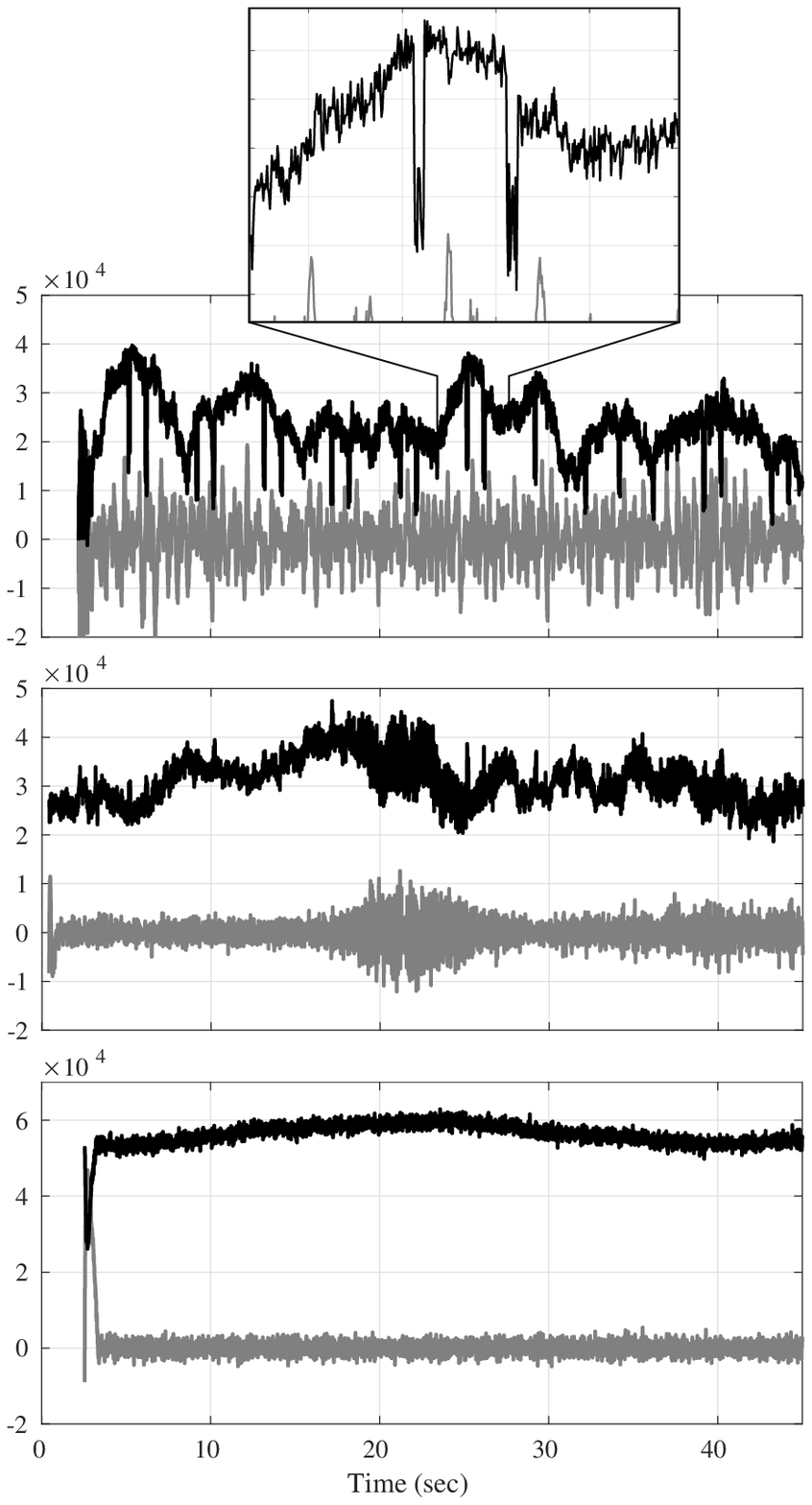}
  \caption{In-phase (black) and quadrature (gray) 10-ms accumulation time
    histories for the strongest false signal from the day 74 capture (top),
    the strongest authentic signal from the day 74 capture (middle), and the
    strongest signal from the day 158 nominal capture (bottom).  Data wipe-off
    was implemented in the middle and bottom figures.  The inset on
    the top panel shows an amplified view of two sudden amplitude fades in the
    received false signal.  The maximum carrier-to-noise ratio $\cnr$ over the
    intervals shown are, from the top, 42.5, 46.8, and 52.5 dB-Hz.}
  \label{fig:iq_plots}
\end{figure}

\subsection{Power Spectral Characteristics}
Figs. \ref{fig:power_spectra} and \ref{fig:bump_wax_and_wane} illustrate the
captured signals' spectral characteristics.  The spectra of narrowband
interference near L2 are simple and remain similar across all three days, but
the wideband interference at L1 is more complex and variable.  It is clear from
the left column of Fig. \ref{fig:power_spectra} that the matched-code
interference is cluttered by other components.  Were it generated by a
high-quality signal simulator, L1 interference would tend to be smooth like the
authentic signals underlying the spectrum shown in the lower left panel.
Instead, it appears to be an amalgam of components.
Fig. \ref{fig:bump_wax_and_wane} reveals that the rounded prominence in the L1
Day 144 panel exhibits oscillatory behavior with a 5-second period.  Whether
such variations are deliberate or are caused by transmitter idiosyncrasies is
unknown.

\subsection{Baseband Signal Characteristics}
Fig. \ref{fig:iq_plots} shows time histories of 10-ms-accumulated complex
correlation products from a false (top panel) and two authentic (bottom two
panels) GPS L1 C/A signals present in the captured L1 band.  The false signal's
empirical $\cnr$ value is 42.5 dB-Hz on average, but the signal is highly
irregular, manifesting both gradual and sudden fading.  The gradual fading may
be a result of scintillation as the signal passes upward through the lower
ionosphere \cite{t_humphreys08_cpt}, but the sudden fading, highlighted in the
inset of the top panel, is unnatural and likely originates at the transmitter.

\subsection{Source Geolocation}
\begin{figure}
  \includegraphics[width=0.5\textwidth]{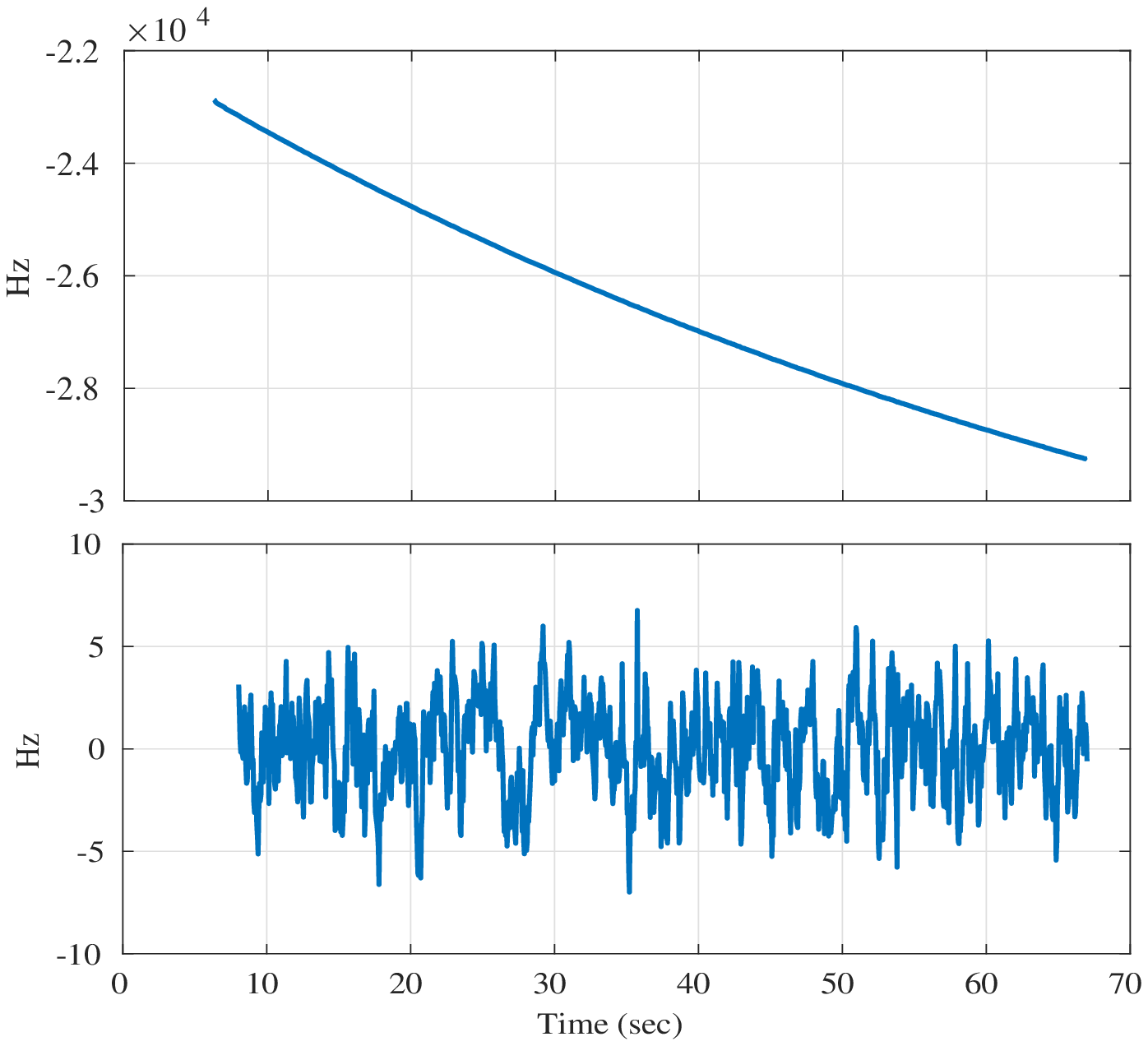}
  \caption{Top: Doppler time history corresponding to the false PRN 10 signal
    from the day 144 capture.  Bottom: Post-fit residuals of the Doppler time
    history assuming the estimated transmitter location and clock rate
    offset.  The standard deviation of the post-fit residuals is 2.3 Hz.}
  \label{fig:fD_and_residuals}
\end{figure}

\begin{figure}
  \includegraphics[width=0.5\textwidth]{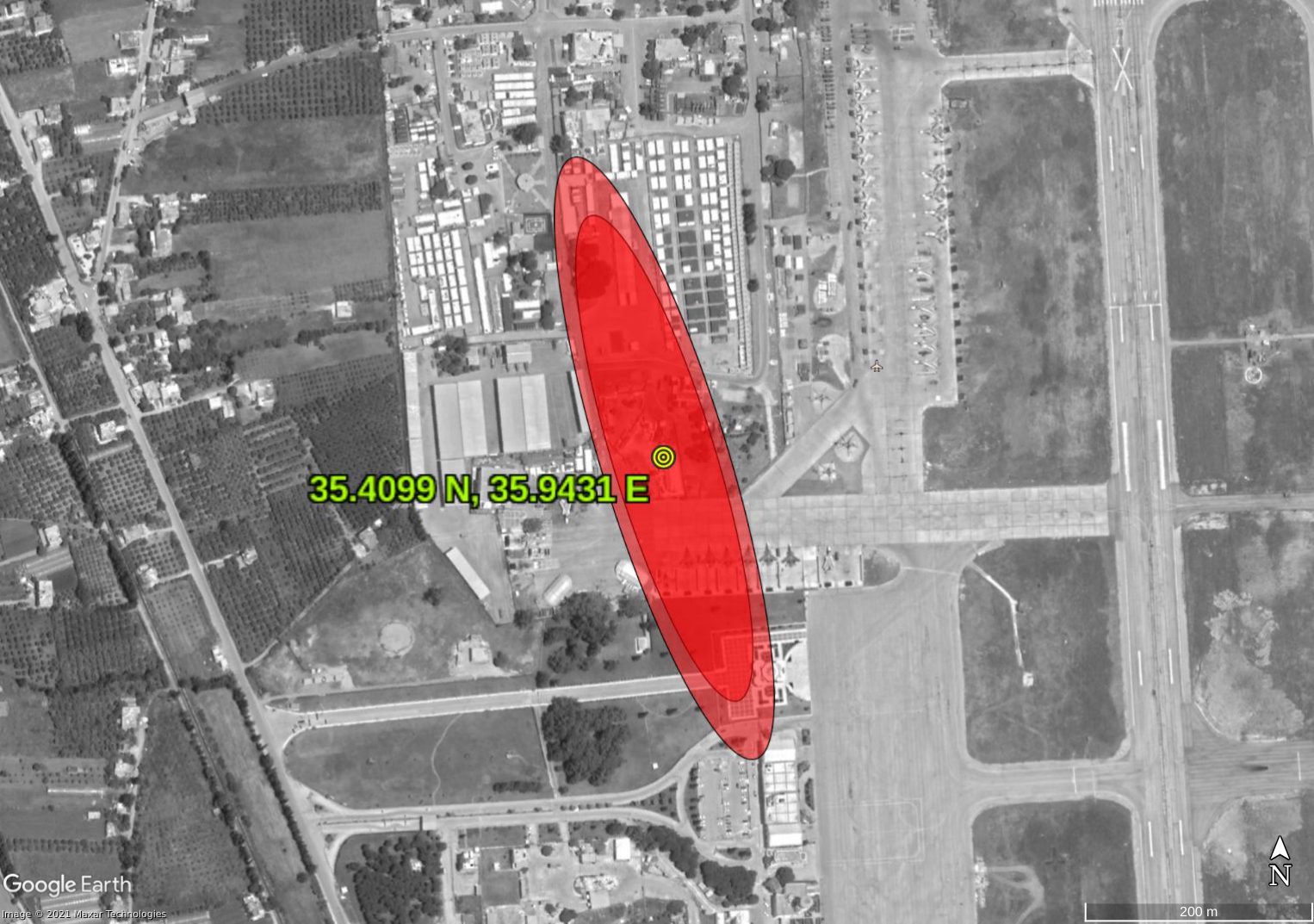}
  \caption{Estimated transmitter location overlaid on formal-error 95\% and
    99\% horizontal error ellipses.  The location is coincident with an airbase
    on the coast of Syria.  The semi-major axis of the 95\% ellipse is 220
    meters.}
  \label{fig:transmitter_location}
\end{figure}

The presence of a trackable carrier signal after despreading (cf. top panel of
Fig. \ref{fig:iq_plots}) enables geolocation of the interference
source as described in Section \ref{sec:single-satell-geol}.  A receiver
navigation solution was first estimated on days 74, 144, and 151 of 2018 using
an Extended Kalman Filter (EKF) drawing in pseudorange and Doppler measurements
extracted from the authentic GPS L1 C/A, GPS L2C, and Galileo E1
signals. Propagation of the receiver state estimate between measurement updates
was based on a nearly-constant acceleration dynamics model.  Time histories of
the quantities $\vb{v}_{\rm R}$, $\delta \dot{t}_{\rm R}$, and the receiver
position component of $\vbh{r}$ were then extracted from the EKF's state
estimate and treated as known for purposes of source geolocation.

A batch estimator for interference source position and clock frequency bias was
formulated as described in Section \ref{sec:single-satell-geol}.  It was
assumed that the interference observed on all three days originated from the
same stationary transmitter, which allowed multiple days of Doppler
measurements, collected on non-repeating ground-tracks, to be combined to form
a tightly-constrained estimate. If these assumptions were false, large post-fit
measurement residuals could be expected to manifest, which was not the case.
Consistent with the assumption of a stationary transmitter, transmitter
altitude was assumed to be near ground-level and was included as a
pseudo-measurement.

A constant transmitter clock frequency error $\delta \dot{t}_{\rm T}$ was
assumed to apply during each capture, but a new value of
$\delta \dot{t}_{\rm T}$ was estimated for each of the three
captures. Comparing the batch-estimator-produced estimates of
$\delta \dot{t}_{\rm T}$ for days 74 and 144 revealed a two-sample transmitter
clock frequency stability of approximately
$\sigma_y(2,T,\tau) = 6.85 \times 10^{-9}$ at a sampling interval $T$ of 70
days and an observation time (averaging interval) $\tau$ of approximately 50
seconds.  The $B_2$ bias function \cite{barnes1969nbs} was used to convert
this two-sample deviation to an Allan deviation, where
$B_2(r, \mu) = 1.8144 \times 10^{5}$ for $r = T/\tau$ and $\mu = 1$, which
assumes $h_{-2}$ is the dominant spectral component.  This yielded an
equivalent Allan deviation for $\tau = 50~\mathrm{seconds}$ of
$\sigma_y(2,\tau,\tau) = 1.6 \times 10^{-11}$, which is consistent with a
standard-quality OCXO \cite{bagala2016improvement}.

Thus, given the results of Table
\ref{table:clock_quality_vs_geolocation_sigma}, treating
$\delta \dot{t}_{\rm T}$ as constant over each 60-second capture can be
conservatively expected to introduce 95\% errors smaller than 720 meters (that
corresponding to a low-quality OCXO) in single-pass geolocation.  A Monte-Carlo
simulation like the one that produced the Table
\ref{table:clock_quality_vs_geolocation_sigma} data but for the combined three
days of collection showed that, assuming independence in the clock frequency
errors between passes, and conservatively assuming a low-quality OCXO, this
error source can be expected to contribute 95\% errors below 230 meters in the
combined 3-day solution.

It is worth noting that, because $\delta \dot{t}_{\rm R}$ and
$\delta \dot{t}_{\rm T}$ enter equivalently into the Doppler measurement model
(\ref{eq:doppler_meas_model}), and because no prior knowledge of these
parameters is assumed in the batch maximum-likelihood estimator, an error in
the EKF's estimate of $\delta \dot{t}_{\rm R}$ will directly manifest in the
batch-estimator-produced estimate of $\delta \dot{t}_{\rm T}$ for each capture.
However, examination of the the EKF's error covariance revealed that its
estimate of $\delta \dot{t}_{\rm R}$ was good to better than
$7 \times 10^{-10}$ (1-$\sigma$) for the day 74 and 144 captures.  Thus,
receiver-side errors are likely small enough that
$\sigma_y(2,\tau,\tau) = 1.6 \times 10^{-11}$ remains an accurate assessment
of the transmitter clock stability.

Fig.~\ref{fig:fD_and_residuals} shows time histories of Doppler and post-fit
residuals for false PRN 10 collected on day 144. The standard deviation of the
post-fit residuals is 2.3 Hz, indicating that the measurement model in
(\ref{eq:doppler_meas_model}), and the assumption of a constant
$\delta \dot{t}_{\rm T}$ over each capture, are reasonably
accurate. Fig.~\ref{fig:transmitter_location} shows the estimated position of
the interference source.  The horizontal error ellipses, which indicate a
solution better than 220 meters (95\%), are formal error ellipses assuming (1)
constant $\delta \dot{t}_{\rm T}$ over each capture, (2) a standard deviation
of 5 m for the transmitter altitude constraint, and (3) a standard deviation
between 2.3 and 2.5 Hz (depending on the empirical post-fit residuals for each
capture) for the measurement error $w$ from (\ref{eq:doppler_meas_model}).
Assuming an OCXO-quality clock in the transmitter, the error caused by modeling
a constant $\delta \dot{t}_{\rm T}$ is small compared to these formal error
ellipses.  While the true location is not known, the geolocation solution based
on the model plausibly coincides with a Russian-operated air base in Syria.

\subsection{Transmitter Power}
In the presence of interference, $\cnr$ actually measures the
carrier-to-interference-and-noise ratio, CINR.  By analyzing the authentic
signal CINR values in the captured data one can infer the transmitted power in
the direction toward ISS of the emitter located in Syria.  The data presented
here are for the day 74 capture.  The average decrease in CINR observed at the
ISS when 1340 km from the source was approximately 6 dB.  One may assume the
interference acts as multi-access interference, whose spectral density is
$I_0 = (2/3) P_{\rm I} T_C$ \cite{humphreysGNSShandbook}, where $P_{\rm I}$ is
the received interference power and $T_C = 1023^{-1}$ ms is the GPS L1 C/A
spreading code chip interval.  Then, assuming $N_0 = -204$ dBW/Hz, a drop in
CINR by 6 dB implies $P_{\rm I} = -137$ dBW.  Let
$P_{\rm S} = P_{\rm I} - G_r + L$, and assume path loss $L = 159$ dB,
consistent with a stand-off distance of 1340 km, and receiver antenna gain
$G_r = 3$ dB.  It follows that the transmitter power of the interference source
in the direction toward the ISS during the day 74 capture is $P_{\rm S} = 19$
dBW, or 79 W.  If the transmitter is focused on ground based targets, then it
is possible that the gain pattern is toroidal.  The elevation angle of the ISS
as seen from the transmitter is low during this period (varying between 8 and
13.5 degrees) and may have been near the maximum of a toroidal gain pattern.

\section{Global Interference Survey via Receiver-Reported CINR}
\label{sec:global-interference-survey}
The raw IF data captures from the ISS FOTON receiver enable detailed monitoring
of GNSS interference signals and their structure, but such captures are
infrequent and limited to short 1-minute intervals. By contrast, the 1-Hz
standard GNSS observables and 100-Hz data-wiped complex correlation products
have been logged nearly continuously since early 2017. These data facilitate a
world-wide survey of strong GNSS interference.

\subsection{Calculation of Receiver-Reported CINR}
Receiver-reported CINR is calculated as
\begin{equation}
  \label{eq:CN0calc}
  \mathrm{CINR} = \left(\frac{\E{I^2 + Q^2}}{2\sigma_{IQ}^2} - 1\right)\frac{1}{T_a}
\end{equation}
where the expectation $\E{I^2 + Q^2}$ is estimated by moving average using a
Euler approximation to a standard low-pass filter
\begin{equation}
\E{I^2 + Q^2}_k = \E{I^2 + Q^2}_{k-1} + K \left(I_k^2 + Q_k^2 - \E{I^2 + Q^2}_{k-1}\right)
\end{equation}
with subscripts $k$ and $k-1$ indicating the current and previous accumulation
interval. The gain parameter $K = \frac{T_a}{\tau}$ with accumulation interval
$T_a = 10$ msec and filter time constant $\tau = 0.5$. $I_k$ and $Q_k$ are the
in-phase and quadrature prompt correlation products for the current
accumulation interval.

The receiver noise floor, $2\sigma_{IQ}^2$, can be derived analytically for a
2-bit quantizing RF front-end and a software-defined GNSS receiver based on
the quantization models of both the RF front-end and receiver local carrier
replica. It can be shown that
\begin{equation}
  2\sigma_{IQ}^2 = 2N\left(a_0^2 p_{a0} + a_1^2 p_{a1}\right)\left(b_0^2 p_{b0} + b_1^2 p_{b1}\right)
\end{equation}
where $N$ is the number of samples per accumulation interval, $a_0$, $a_1$,
$b_0$, and $b_1$ are the low and high quantization values of the RF front-end
and local carrier replica, respectively, and $p_{a0}$, $p_{a1}$, $p_{b0}$, and
$p_{b1}$ are their associated probabilities. In practice, $p_{a0}$ and $p_{a1}$
depend on the implementation of the automatic gain control (AGC) in
the RF front end, and $p_{b0}$ and $p_{b1}$ are selected by the receiver
designer, e.g., to minimize quantization distortion.  The following values are
applicable to the FOTON receiver
\begin{align}
  a_0 & = 1, a_1 = 3, b_0 = 1, b_1 = 3 \nonumber \\
  p_{a0} & = 0.68269, p_{a1} = 0.31731, p_{b0} = 0.38418, p_{b1} = 0.61582
  \nonumber \\
  N & = 5714.286 \nonumber
\end{align}
which yield a noise floor of $2\sigma_{IQ}^2 = 239669$ front-end units.

\subsection{Methodology}
The carrier power $C$ of an authentic signal can be modeled as a function
$C(j, f, r_{sr}, z_s, z_r)$, where $j$ is the GNSS satellite identifier (SV
ID), $f$ is the frequency band (L1 or L2), $r_{sr}$ is the range between the
GNSS satellite antenna and the ISS FOTON antenna, $z_s$ is the angle between
the satellite boresight direction and the direction to the ISS antenna (i.e.,
the satellite antenna off-boresight angle), and $z_r$ is the angle between the
ISS antenna boresight direction and the direction to the satellite (receiver
antenna off-boresight angle). A hypothesis test based on the receiver-reported
CINR can be designed to detect whether ($H_1$) or not ($H_0$) the receiver is
experiencing interference. Under a given $P_\mathrm{F}$, this requires that the
statistics $\mathbb{E}[l|H_0]$ and $\mathrm{Var}(l|H_0)$ be known. To obtain
these statistics, this section assumes the receiver reports interference-free
data (consistent with $H_0$) when the ISS is over deep ocean bodies.

\begin{figure}
  \centering \includegraphics[width=0.4\textwidth]{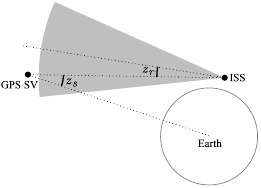}
  \caption{For receiver off-boresight angle $z_r \leq 15^\circ$ (within the
    gray region), the satellite off-boresight angle $z_s$ is restricted between
    $14.2^\circ \leq z_s \leq 15.2^\circ$}
  \label{fig:geometry}
\end{figure}

To isolate the variations in reported CINR due to interference, the data are
first pre-processed to eliminate the predictable sources of carrier power
variation. First, the dependence of $C$ on $r_{sr}$ is removed by compensating
for the free space path loss:
\[
  \hat{C}(j, f, z_s, z_r) = C(j, f, r_{sr}, z_s, z_r) \times {\left( \frac{4 \pi r_{sr} f}{c} \right)}^2
\]
Modeling of interference-free $\cnr$ is complicated by the ISS's local
multipath environment. The ISS antenna is flanked by solar panels that move
with respect to the FOTON antenna, causing a non-stationary signal obstruction
and multipath environment. Nevertheless, an off-boresight angle window
$z_r \in [0^\circ, 15^\circ]$ is known to be free of obstructions.  Only the
signals received in this window are considered for interference detection in
this paper's analysis. Confining $z_r$ to this window restricts the geometry
between GNSS satellites and the ISS such that
$z_s \in [14.2^\circ, 15.2^\circ]$ (see Fig.~\ref{fig:geometry}). The GNSS
antenna gain pattern can be assumed to be relatively constant over
$\pm$0.5$^\circ$.  Thus, $\hat{C}(j, f, z_s, z_r)$ can be assumed independent
of $z_s$.

The mean and variance of ISS-reported range-compensated-CINR values
$\hat{C}/N_0$ collected over deep ocean regions are maintained as control data
in a three-dimensional grid of SV ID $j$, frequency band $f$, and receiver
off-boresight angle $z_r$.  For a world-wide analysis of GNSS interference
events, a hypothesis test is performed on the test statistic derived from
$\hat{C}/N_0$ values that fall within $z_r \in [0^\circ, 15^\circ]$. The test
is performed separately for the L1 and L2 bands since the interference
characteristics are frequency dependent. If the reported test statistics falls
below $\mathbb{E}[l|H_0] - 3 \sqrt{\mathrm{Var}(l|H_0)}$, the receiver is
declared to be under interference.  This threshold respects a $P_\mathrm{F}$ of
approximately $1.35 \times 10^{-3}$.

\subsection{Discussion of Results}
\begin{figure*}[!ht]
  \centering
  \begin{subfigure}[b]{\textwidth}
    \includegraphics[width=\textwidth, trim=65 345 70 321, clip]{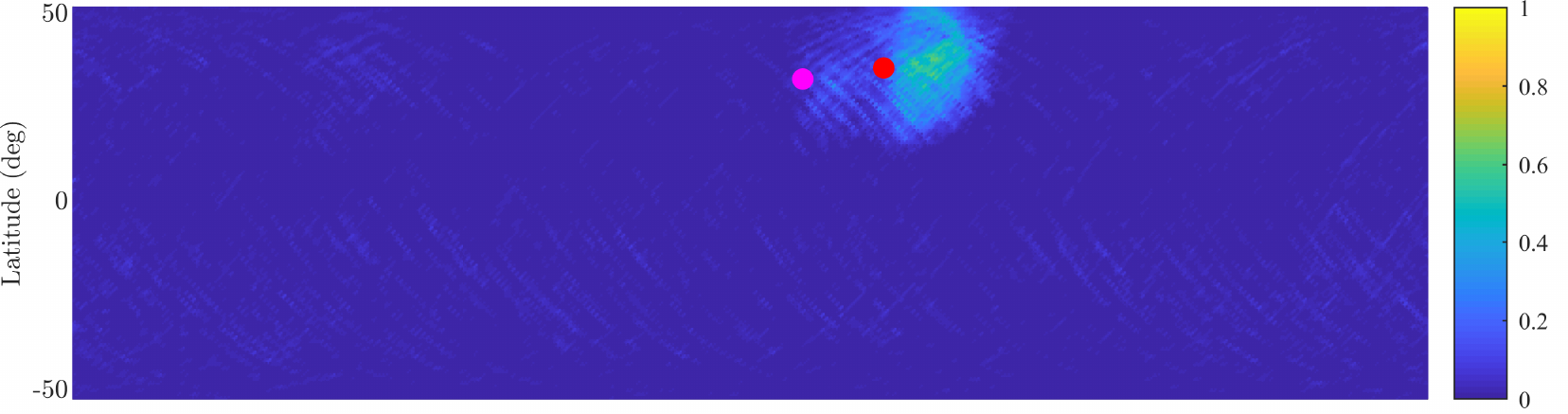}
  \end{subfigure}
  \begin{subfigure}[b]{\textwidth}
    \includegraphics[width=\textwidth, trim=65 330 70 321, clip]{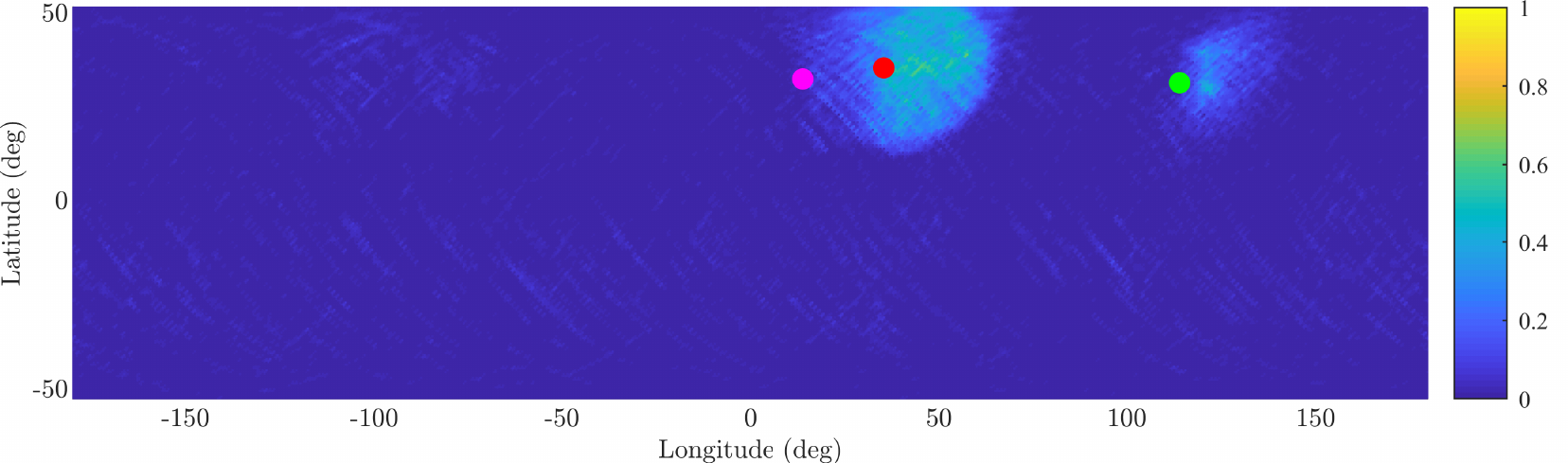}
  \end{subfigure}
  \caption{Ratio of number of potential GPS L1 (top panel) and L2 (bottom
    panel) interference events recorded to total number of hypothesis tests
    performed at each location on the map for the full span of data considered
    in this paper, from March 2017 to June 2020.  The red dots indicate the
    estimated origin of the interference from Syria based on raw IF
    recordings. Another hotspot of interference is apparent to the west of the
    Syrian location. The magenta dots denote the approximate location of
    GNSS interference reports in the Libyan region~\cite{uscgstatus}. In
    addition to the interference over the Syrian and Libyan regions, strong L2
    interference over mainland China is observed. The green dot at
    (\SI{32}{\degree} N, \SI{114}{\degree} E) indicates a hypothesized
    interference source location based on the shape and location of the
    observed hotspot.}
  \label{fig:fL1_and_fL2_violations}
\end{figure*}

Fig.~\ref{fig:fL1_and_fL2_violations} shows the ratio of the number of
potential interference events recorded at L1 (top panel) and L2 (bottom panel)
to total number of hypothesis tests performed at each location for the
foregoing detection threshold.  As expected, a high ratio of potential
interference events is reported for both L1 and L2 near Syria (marked with a
red dot). Note that the interference ``hotspot'' appears to the east of the
source because the ISS orbit is prograde and the FOTON antenna points in the
anti-velocity direction. In other words, the FOTON antenna is exposed to
interference only after the ISS passes eastward over an emitter's location.

The high values of the statistic for both L1 and L2 east of Syria indicate that
the interference activity in Syria has been persistent over nearly the full
interval considered in this paper, from March 2017 to June 2020.  A monthly
analysis (not shown) revealed that the source has been transmitting at L2 since
no later than March 2017.  It began transmitting weak interference at L1 during
the second half of 2017, then much stronger interference at L1 during the first
quarter of 2018.  The interference at L1 and L2 was ongoing in June 2020.

An additional hotspot is present to the west of the Syrian location. This
hotspot, which emerged in the second half of 2019, is consistent with reports
of GNSS interference in the Libyan region~\cite{uscgstatus}. The magenta dots
in Fig.~\ref{fig:fL1_and_fL2_violations} denote the approximate location of the
area in which interference has been documented (\SI{33}{\degree} N,
\SI{14}{\degree} E).  Fig.~\ref{fig:fL1_and_fL2_violations} also reveals strong
L2 interference over mainland China.  This interference has been present since
March 2017 at the latest and was ongoing in June 2020. The green dot in
Fig.~\ref{fig:fL1_and_fL2_violations}, marked at (\SI{32}{\degree} N,
\SI{114}{\degree} E), indicates a hypothesized interference source location
based on the shape and location of the observed hotspot.

Note that the above method of counting potential interference events based on
CINR degradation ignores cases where interference might lead to complete loss
of track of some or all GPS signals. However, the data from the ISS shows that
FOTON does not lose track of authentic GNSS signals even when flying by the
strong interference source in Syria. In fact, the reported CINR over Syria is
well above the weakest signal that FOTON is capable of tracking. As a result,
it was concluded that in cases where FOTON seems to track few or no GPS
signals, it is likely due to some abnormal behavior of the receiver, and not
due to a potential interference event.

\begin{figure}
  \centering
  \includegraphics[width=0.475\textwidth]{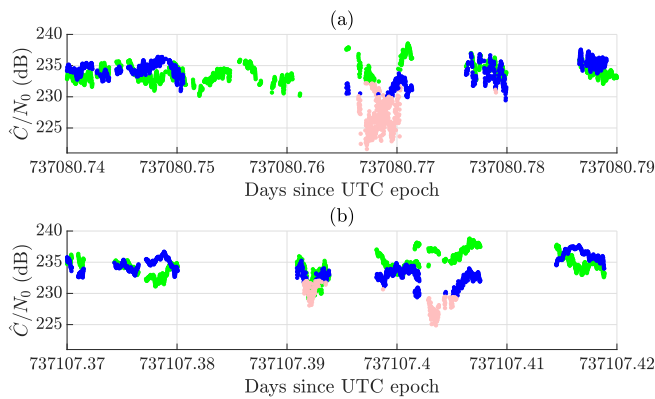}
  \caption{Time histories of range-compensated receiver-reported CINR as the
  ISS flies over potential GPS interference zones over Syria and China.}
  \label{fig:chatn0}
\end{figure}

In addition to the global average analysis summarized in
Fig.~\ref{fig:fL1_and_fL2_violations}, it is instructive to examine the time
history of receiver reported CINR as the ISS passes over an interference
hotspot.  Fig.~\ref{fig:chatn0} shows two such histories for signals within the
admissible range of $z_r$ as the ISS goes over the strong interference regions
in Syria (Fig.~\ref{fig:chatn0}(a)) and China (Fig.~\ref{fig:chatn0}(b)). Green
and blue data points represent range-compensated CINR values for authentic L1
and L2 GNSS signals, respectively, above the applicable threshold, which
depends on $i$, $f$, and $z_r$.  Light red data points are the same data when
below the applicable threshold.  Both L1 and L2 signals are declared under
interference in Fig.~\ref{fig:chatn0}(a), whereas only L2 signals are declared
under interference in Fig.~\ref{fig:chatn0}(b). The brief dip in
Fig.~\ref{fig:chatn0}(b) prior to the major dip over China is caused by the
interference originating in Syria. Gaps in the time histories indicate periods
with no tracked signals in the admissible off-boresight angle window.

\section{Implications for GNSS Receivers}
\label{sec:impl-gnss-rece}
The matched-code interference captured over Syria is intriguing.  So far as
this paper's authors are aware, no other GNSS interference captured from an
operational (as opposed to experimental) source has exhibited the
characteristics observed in the interference emanating from Syria.  If the
intent behind the signals transmitted at L1 is not spoofing but rather denial
of GPS service, as might be inferred from the lack of navigation data bit
modulation, then it is unclear why an ensemble of signals, each one modulated
by a separate GPS L1 C/A spreading code, was transmitted. The transmitter could
just as well allocate its power to a single GPS L1 C/A spreading code, or any
code with a similar spectral density.  However, transmitting a multitude of
spreading codes can be effective at disrupting cold-start acquisition of GPS L1
C/A signals, as explained below.

\subsection{Efficient Jamming}
The art of jamming is more sophisticated than merely emitting RF energy into a
target band.  An efficient jammer is one that effectively disrupts GNSS
service in a given area of operations but does so with as little power as
possible.  Such frugality extends the life of battery-powered jammers, and
makes jammers less conspicuous.  The key to efficient jamming is avoiding
wasteful allocation of signal power.  Obviously, allocating power outside a
target receiver's passband is wasteful because the interference is filtered out
by the receiver's RF front-end.  Less obviously, narrowband jamming applied
directly in the passband is also wasteful.  To understand this, consider the
vector space of all possible input signals, and a partitioning into a subspace
that contains the jamming signal and one that does not.  If the jammer-occupied
subspace is sparse with respect to the desired signal subspace, and if the
receiver's front-end amplification and quantization are not saturated, then a
technique can be developed to excise the jammer-occupied subspace with minimal
degradation to the desired signals.  For a narrowband jammer, the technique is
notch filtering; for a pulsed jammer, the technique is pulse blanking
\cite{humphreysGNSShandbook}.

An efficient jammer maximizes overlap with the desired-signal subspace for a
given power allocation.  Jamming that is continuous in the time domain and
white (spectrally flat) within the desired signal passband in the frequency
domain is reasonably efficient because it extensively overlaps the desired
signal subspace.  Continuous-time matched-spectrum jamming is even more
efficient: Instead of spreading the jamming power evenly across the passband, a
matched-spectrum jammer shapes it for greater overlap with the desired signal
subspace.  Consider a random binary spreading code with chip interval
$T_{\rm C}$.  Suppose a spectrally-flat jammer is designed to cover the
spreading code's primary spectral lobe and first two side lobes, for a total
frequency span of $4/T_{\rm C}$ Hz.  The noise power density that passes
through the receiver's matched filter is $I_0 = P_{\rm I} T_{\rm C}/4$, where
$P_{\rm I}$ is the interference power.  By contrast, for a matched-spectrum
jammer $I_0 = (2/3)P_{\rm I} T_{\rm C}$ \cite{humphreysGNSShandbook}.  When
$I_0$ is large enough that CINR $\approx C/I_0$, the matched-spectrum jammer is
4.3 dB more potent than the spectrally-flat jammer.  What is more, the
spectrally-flat jammer spanning $4/T_{\rm C}$ Hz can be excised by filtering in
the frequency domain: even if the main lobe and adjacent two side lobes of the
authentic signals are removed along with the jamming, the authentic signals are
only attenuated by 13 dB.  The spectrally-flat jammer must spread its power
even wider to avoid such excision by filtering, resulting in an even less
favorable potency compared with matched-spectrum jamming.  By contrast, a
matched-spectrum jammer cannot be excised by filtering because its spectrum
follows the $\sinc^2(fT_{\rm C})$ envelope of the authentic
binary-code-modulated signals.  Thus, spectrum matching is a necessary
condition for efficient jamming.

However, spectrum matching is not a sufficient condition for effective jamming.
Consider a jammer emitting a carrier modulated only by a single publicly-known
spreading code of arbitrary length.  This signal is sparse with respect to the
desired signal subspace.  It can be excised by the receiver generating a local
replica of the interference signal, aligning this replica's code phase, carrier
phase, and amplitude with the interference signal, and subtracting the replica
from the digitized output of the receiver's RF front end.  Assuming sufficient
front-end bit depth and amplifier linearity, this procedure can be extended to
an arbitrary number of such interference signals, each with a known waveform;
the technique is known as successive interference cancellation (SIC)
\cite{madhani03_sic}.

Thus, an effective jammer will avoid predictable signals: a more sophisticated
approach to spectrum matching is modulation of the carrier with a non-repeating
spectrum-matching spreading code known only to the jammer.  But this is only
necessary when the target receiver is capable of SIC.  If, for example, the
receiver has no way of distinguishing authentic from interference signals, then
it cannot apply SIC without also eliminating desired signals.

\subsection{The Cold Start Vulnerability}
It is instructive to consider the conditions under which a GNSS receiver is
unable to distinguish between authentic and interference signals: (1) the
authentic and interference signals are identical in all aspects of significance
(modulation, code phase, carrier phase and frequency, amplitude), or (2) the
authentic and interference signals are identical except in ways the target
receiver is unable to exploit to distinguish them.  In case (1), the
interference is hardly a problem: it simply reinforces the authentic signals.
Case (2) is more interesting.  Let the term \emph{spoofing interference} refer
to matched-code interference with all additional modulation requisite to make
the interference signal's structure and content identical to an authentic
signal's.  If a receiver is exposed to spoofing interference while already
tracking enough authentic signals to form a navigation solution and when in
possession of accurate satellite ephemerides, it can distinguish any authentic
and interference signals that differ in code phase, carrier frequency, or
amplitude. (It can additionally distinguish by carrier phase if performing
precise carrier-based navigation.)  Therefore, jamming a navigation-locked
receiver with spoofing interference may be ineffective because the target
receiver can apply SIC.

However, during a cold start, the target receiver's time and position are
uncertain, and it lacks the ephemerides necessary to predict the code phase and
Doppler of authentic signals even if its time and position were known.  In this
case the receiver is highly vulnerable to spoofing interference.  Suppose a
jammer generates a counterpart power-matched spoofing signal for each authentic
GNSS signal available in an area of operations.  Suppose further that the
ensemble of spoofing signals is self-consistent with a location and time
different from the target receiver's true location and time.  On cold start,
the receiver is jammed not in the traditional sense of being unable to acquire
and track the authentic signals, but rather in the sense of being unable to
confidently declare which of two plausible-looking navigation solutions is
correct.  If, under this circumstance, the receiver refuses to provide a
navigation solution, the user is effectively denied GNSS service.  If instead
the receiver mistakenly provides the spoofed solution, the user could be
exposed to hazardously misleading information.

This type of spoofing interference is highly efficient.  Suppose the target
receiver has a cold-start CINR acquisition threshold of $\eta$ dB-Hz.  Then
traditional matched-spectrum jamming would require a jamming-to-authentic power
ratio equal to
\begin{equation}
  \label{eq:denyTrad}
  \frac{P_{\rm I}}{C} = -\left[\eta + 10\log_{10}\left(\frac{2 T_{\rm C}}{3} \right)\right]
\end{equation}
which, for GPS L1 C/A signals and a typical $\eta = 30$ dB-Hz, amounts to
$31.8$ dB.  By contrast, jamming via single-counterpart power-matched spoofing
interference requires only $P_{\rm I}/C = 0$ dB, which makes it more
than 1500 times more efficient for denial of GNSS service at cold start.

\subsection{Discussion}
The interference captured over Syria causes traditional matched-spectrum
jamming at close range, and is capable of disrupting cold-start acquisition far
beyond this (along its line-of-sight).  Indeed, it would be at least partially
effective at preventing FOTON cold start even at the maximum line-of-sight
range to the ISS, or approximately 1600 km.  However, the interference signals
as broadcast have at least four flaws, any one of which could be exploited by
receivers to distinguish them from authentic signals: (1) they lack navigation
data modulation; (2) they are broadcast on a common and constant carrier
frequency; (3) they share a common code phase alignment; (4) they include
signals for (almost) all GPS PRNs.  A receiver built to detect these anomalies
could identify the imposter signals and eliminate them via SIC.

However, in general, spoofing interference is not so easily distinguished from
authentic signals, and can be both effective and power-efficient at denying
GNSS service on cold start.  The best defense against spoofing interference
intended to deny GNSS service remains an open problem.

\section{Conclusions}
Low-earth-orbiting instruments capable of receiving signals in GNSS bands are a
powerful tool for characterizing GNSS interference emanating from terrestrial
sources.  Data from one such instrument, the FOTON software-defined GNSS
receiver, which has been operational on the International Space Station since
February 2017, reveal interesting patterns of GNSS interference from March 2017
to June 2020.  Signals from a particularly powerful and persistent interference
source active in Syria since 2017 were captured and characterized, and the
source was geolocated to better than 1 km.  A global analysis revealed other
interference hotspots around the globe in both the GPS L1 and L2 frequency
bands.  Matched-code interference such as emitted at the GPS L1 frequency by
the jammer in Syria is power-efficient for jamming signal-locked GNSS
receivers.  GNSS receivers are particularly vulnerable to such interference
during cold start.

\section*{Acknowledgments}
Work at The University of Texas has been supported in part by the National
Science Foundation under Grant No. 1454474 (CAREER) and in part by in part by
the U.S. Department of Transportation (USDOT) under Grant 69A3552047138 for the
CARMEN University Transportation Center (UTC). Work at the Naval Research
Laboratory was supported by the Chief of Naval Research.  The STP-H5/GROUP-C
experiment was integrated and flown under the direction of the Department of
Defense Space Test Program.

\bibliographystyle{IEEEtran}
\bibliography{pangea}

\begin{thebibliography}{10}
\providecommand{\url}[1]{#1}
\csname url@samestyle\endcsname
\providecommand{\newblock}{\relax}
\providecommand{\bibinfo}[2]{#2}
\providecommand{\BIBentrySTDinterwordspacing}{\spaceskip=0pt\relax}
\providecommand{\BIBentryALTinterwordstretchfactor}{4}
\providecommand{\BIBentryALTinterwordspacing}{\spaceskip=\fontdimen2\font plus
\BIBentryALTinterwordstretchfactor\fontdimen3\font minus
  \fontdimen4\font\relax}
\providecommand{\BIBforeignlanguage}[2]{{%
\expandafter\ifx\csname l@#1\endcsname\relax
\typeout{** WARNING: IEEEtran.bst: No hyphenation pattern has been}%
\typeout{** loaded for the language `#1'. Using the pattern for}%
\typeout{** the default language instead.}%
\else
\language=\csname l@#1\endcsname
\fi
#2}}
\providecommand{\BIBdecl}{\relax}
\BIBdecl

\bibitem{lightsey2013demonstration}
E.~G. Lightsey, T.~E. Humphreys, J.~A. Bhatti, A.~J. Joplin, B.~W. O'Hanlon,
  and S.~P. Powell, ``Demonstration of a space capable miniature dual frequency
  {GNSS} receiver,'' \emph{Navigation}, vol.~61, no.~1, pp. 53--64, Mar. 2014.

\bibitem{ao2009rising}
C.~O. Ao, G.~A. Hajj, T.~K. Meehan, D.~Dong, B.~A. Iijima, A.~J. Mannucci, and
  E.~R. Kursinski, ``Rising and setting {GPS} occultations by use of open-loop
  tracking,'' \emph{Journal of Geophysical Research}, vol. 114, no.~D4, Feb.
  2009.

\bibitem{jin2010gnss}
S.~Jin and A.~Komjathy, ``{GNSS} reflectometry and remote sensing: New
  objectives and results,'' \emph{Advances in Space Research}, vol.~46, no.~2,
  pp. 111--117, July 2010.

\bibitem{isoz2014int}
O.~Isoz, S.~A. Buehler, K.~Kinch, M.~Bonnedal, and D.~M. Akos, ``Interference
  from terrestrial sources and its impact on the {GRAS} {GPS} radio occultation
  receiver,'' \emph{Radio Science}, vol.~49, no.~1, pp. 1--6, Jan. 2014.

\bibitem{dempster2016interference}
A.~G. Dempster and E.~Cetin, ``Interference localization for satellite
  navigation systems,'' \emph{Proceedings of the {IEEE}}, vol. 104, no.~6, pp.
  1318--1326, June 2016.

\bibitem{ho1997geolocation}
K.~Ho and Y.~Chan, ``Geolocation of a known altitude object from {TDOA} and
  {FDOA} measurements,'' \emph{{IEEE} Transactions on Aerospace and Electronic
  Systems}, vol.~33, no.~3, pp. 770--783, July 1997.

\bibitem{pattison2000sensitivity}
T.~Pattison and S.~Chou, ``Sensitivity analysis of dual-satellite
  geolocation,'' \emph{{IEEE} Transactions on Aerospace and Electronic
  Systems}, vol.~36, no.~1, pp. 56--71, 2000.

\bibitem{griffin2002interferometric}
C.~Griffin and S.~Duck, ``Interferometric radio-frequency emitter location,''
  \emph{{IEE} Proceedings - Radar, Sonar and Navigation}, vol. 149, no.~3, p.
  153, 2002.

\bibitem{amar2008localization}
A.~Amar and A.~Weiss, ``Localization of narrowband radio emitters based on
  doppler frequency shifts,'' \emph{{IEEE} Transactions on Signal Processing},
  vol.~56, no.~11, pp. 5500--5508, Nov. 2008.

\bibitem{bhatti2015dissertation}
J.~Bhatti, ``Sensor deception detection and radio-frequency emitter
  localization,'' Ph.D. dissertation, The University of Texas at Austin, Aug.
  2015.

\bibitem{smith1989time}
W.~Smith and P.~Steffes, ``Time delay techniques for satellite interference
  location system,'' \emph{{IEEE} Transactions on Aerospace and Electronic
  Systems}, vol.~25, no.~2, pp. 224--231, Mar. 1989.

\bibitem{ho1993solution}
K.~Ho and Y.~Chan, ``Solution and performance analysis of geolocation by
  {TDOA},'' \emph{{IEEE} Transactions on Aerospace and Electronic Systems},
  vol.~29, no.~4, pp. 1311--1322, 1993.

\bibitem{haworth1997interference}
D.~Haworth, N.~Smith, R.~Bardelli, and T.~Clement, ``Interference localization
  for {EUTELSAT} satellites--the first {E}uropean transmitter location
  system,'' \emph{International journal of satellite communications}, vol.~15,
  no.~4, pp. 155--183, 1997.

\bibitem{kalantari2016frequency}
A.~Kalantari, S.~Maleki, S.~Chatzinotas, and B.~Ottersten, ``Frequency of
  arrival-based interference localization using a single satellite,'' in
  \emph{2016 8th Advanced Satellite Multimedia Systems Conference and the 14th
  Signal Processing for Space Communications Workshop ({ASMS}/{SPSC})}.\hskip
  1em plus 0.5em minus 0.4em\relax {IEEE}, Sept. 2016.

\bibitem{murrian2019leoIon}
M.~J. Murrian, L.~Narula, and T.~E. Humphreys, ``Characterizing terrestrial
  {GNSS} interference from low earth orbit,'' in \emph{Proceedings of the {ION}
  {GNSS}+ Meeting}.\hskip 1em plus 0.5em minus 0.4em\relax Institute of
  Navigation, Oct. 2019.

\bibitem{becker1992efficient}
K.~Becker, ``An efficient method of passive emitter location,'' \emph{{IEEE}
  Transactions on Aerospace and Electronic Systems}, vol.~28, no.~4, pp.
  1091--1104, 1992.

\bibitem{ellis2020use}
P.~Ellis, D.~V. Rheeden, and F.~Dowla, ``Use of doppler and doppler rate for
  {RF} geolocation using a single {LEO} satellite,'' \emph{{IEEE} Access},
  vol.~8, pp. 12\,907--12\,920, 2020.

\bibitem{brown2012introKf}
R.~G. Brown and P.~Y. Hwang, \emph{Introduction to Random Signals and Applied
  Kalman Filtering}.\hskip 1em plus 0.5em minus 0.4em\relax Wiley, 2012.

\bibitem{crassidis2011optimal}
J.~L. Crassidis and J.~L. Junkins, \emph{Optimal estimation of dynamic
  systems}.\hskip 1em plus 0.5em minus 0.4em\relax Chapman and Hall/CRC, 2011.

\bibitem{humphreys2019deepUrbanIts}
T.~E. Humphreys, M.~J. Murrian, and L.~Narula, ``Deep-urban unaided precise
  global navigation satellite system vehicle positioning,'' \emph{{IEEE}
  Intelligent Transportation Systems Magazine}, vol.~12, no.~3, pp. 109--122,
  2020.

\bibitem{humphreysGNSShandbook}
T.~E. Humphreys, \emph{Interference}.\hskip 1em plus 0.5em minus 0.4em\relax
  Springer International Publishing, 2017, pp. 469--503.

\bibitem{psiakiNewBlueBookspoofing}
M.~L. Psiaki and T.~E. Humphreys, \emph{Position, Navigation, and Timing
  Technologies in the 21st Century: Integrated Satellite Navigation, Sensor
  Systems, and Civil Applications}.\hskip 1em plus 0.5em minus 0.4em\relax
  Wiley-IEEE, 2020, vol.~1, ch. {C}ivilian {GNSS} {S}poofing, {D}etection, and
  {R}ecovery, pp. 655--680.

\bibitem{welch1967use}
P.~Welch, ``The use of fast fourier transform for the estimation of power
  spectra: A method based on time averaging over short, modified
  periodograms,'' \emph{{IEEE} Transactions on Audio and Electroacoustics},
  vol.~15, no.~2, pp. 70--73, June 1967.

\bibitem{t_humphreys08_cpt}
T.~E. Humphreys, M.~L. Psiaki, and P.~M. Kintner, ``Modeling the effects of
  ionospheric scintillation on {GPS} carrier phase tracking,'' \emph{{IEEE}
  Transactions on Aerospace and Electronic Systems}, vol.~46, no.~4, pp.
  1624--1637, Oct. 2010.

\bibitem{barnes1969nbs}
J.~A. Barnes, \emph{Tables of bias functions, B$_1$ and B$_2$, for variances
  based on finite samples of processes with power law spectral
  densities}.\hskip 1em plus 0.5em minus 0.4em\relax National Bureau of
  Standards, 1969.

\bibitem{bagala2016improvement}
T.~Bagala, A.~Fibich, P.~Kubinec, and V.~Stofanik, ``Improvement of short-term
  frequency stability of the chip scale atomic clock,'' in \emph{2016 {IEEE}
  International Frequency Control Symposium ({IFCS})}.\hskip 1em plus 0.5em
  minus 0.4em\relax {IEEE}, may 2016.

\bibitem{uscgstatus}
{{United States Coast Guard}}, ``{GPS} problem reports status,''
  \url{https://navcen.uscg.gov/?Do=gpsreportstatus}, accessed 2020-08-31.

\bibitem{madhani03_sic}
P.~Madhani, P.~Axelrad, K.~Krumvieda, and J.~Thomas, ``Application of
  successive interference cancellation to the {GPS} pseudolite near-far
  problem,'' \emph{{IEEE} Transactions on Aerospace and Electronic Systems},
  vol.~39, no.~2, pp. 481--488, April 2003.

\end{thebibliography}
\end{document}